\documentclass[fleqn,usenatbib]{mnras}

\usepackage{newtxtext,newtxmath}
\usepackage[T1]{fontenc}

\DeclareRobustCommand{\VAN}[3]{#2}
\let\VANthebibliography\thebibliography
\def\thebibliography{\DeclareRobustCommand{\VAN}[3]{##3}\VANthebibliography}

\usepackage{graphicx}	% Including figure files
\usepackage{amsmath}	% Advanced maths commands
\usepackage{hyperref}
\usepackage{threeparttable}

\title[Superdiffusion of cosmic rays in the vicinity of their accelerators and the resulting $\gamma$-ray emission]{Superdiffusion of cosmic rays in the vicinity of their accelerators and the resulting $\gamma$-ray emission}

\author[Z. Shi et al.]{
Zhaodong Shi,$^{1,2}$\thanks{E-mail: shizd@ustc.edu.cn (ZDS)}
Guangwei Wang,$^{2}$\thanks{E-mail: wangguangwei@mail.ustc.edu.cn (GWW)}
Rui-zhi Yang$^{1,2,3}$\thanks{E-mail: yangrz@ustc.edu.cn (RZY)}
\\
$^{1}$Department of Astronomy, School of Physical Sciences, University of Science and Technology of China, Hefei, 230026, Anhui, China\\
$^{2}$School of Astronomy and Space Science, University of Science and Technology of China, Hefei, 230026, Anhui, China\\
$^{3}$TIANFU Cosmic Ray Research Center, Chengdu, Sichuan, China
}

\date{Accepted XXX. Received YYY; in original form ZZZ}

\pubyear{2026}

\begin{document}
\label{firstpage}
\pagerange{\pageref{firstpage}--\pageref{lastpage}}
\maketitle

\begin{abstract}
We study the distribution of cosmic rays (CRs) in the vicinity of their accelerators, assuming that the transport of CRs in the interstellar medium surrounding the accelerators is described by the superdiffusion, beyond the normal diffusion. We find that the superdiffusivity, which is characterized by the superdiffusion parameter $\alpha$, impacts significantly the distribution of CRs. For impulsive injection, the CR distribution behaves a constant radial profile, except with a power-law tail for $\alpha < 2$ or a Gaussian tail for $\alpha=2$ at large distance, $r$, from the accelerators. For stationary injection, the radial profile of CR protons tends to being proportional to $r^{\alpha - 3}$, while that of CR electrons can deviate from the $r^{\alpha - 3}$ profile, due to their severe energy losses. We also compute the $\gamma$-ray emission, produced by the interactions of CRs with ambient gas and radiation fields, within 100 pc regions around the accelerators. We find that by investigating the $\gamma$-ray morphology, we can distinguish the superdiffusion from the normal diffusion with present and next-generation imaging air Cherenkov telescopes.
\end{abstract}


%%%%%%%%%%%%%%%%%%%%%%%%%%%%%%%%%%%%%%%%%%%%%%%%%%

%%%%%%%%%%%%%%%%% BODY OF PAPER %%%%%%%%%%%%%%%%%%

\section{Introduction} \label{sec:intro}

% importance of CR propagation
Cosmic rays (CRs) are one of the most important ingredients of the interstellar medium (ISM) \citep{Ferriere2001, Grenier2015}. They contribute about 1/4 of the total energy density of the ISM \citep{webber98}, and play an important role in the star forming processes and the dynamics of galaxies \citep{papadopoulos10, breitschwerdt91, Ruszkowski2023}. The transport of CRs in the ISM is conventionally described as a normal diffusion process coupling with the interstellar magnetic fields \citep{Berezinskii1990,Amato2018}. In the standard picture, CRs are scattered via small-scale magnetic irregularities superimposed on the large-scale mean field, which is dubbed as quasi linear theory (QLT) \citep{Jokipii1966, Kulsrud1969, Skilling1971, Schlickeiser1989}. In the normal diffusion, the movement of particles can be described by the Brownian motion \citep{Chandrasekhar1943}, and the mean squared displacement (MSD) of particles scales linearly with time, i.e., $\langle \Delta x^2\rangle \propto t$.

% theoretical motivation of fractional diffusion

However, accumulated observational evidence reveals that the ISM is a turbulent, magnetized and nonuniform multiphase gas, filled with structures on all resolvable spatial scales \citep{Armstrong1995, Ferriere2001, Cox2005} and that the interstellar turbulence is intermittent \citep{Elmegreen2004, Scalo2004, Falgarone2015, Fraternale2020}. Thus the intermittency of turbulence and structures in the ISM can impact significantly the transport of CRs \citep{Kempski2022,Butsky2024,Ewart2026}, and a complete description of CR transport in the ISM can fall beyond the scope of the QLT \citep{Effenberger2025}. Indeed, the compatibility of the standard picture of CR transport with observations has been questioned recently \citep{Gabici2019,Hopkins2022}. The recent advancement in magnetohydrodynamic (MHD) turbulence also brings significant improvement in understanding the microscopic physics related to CR transport in different phases of ISM \citep{Xu2016}. Moreover, the recent theoretical and numerical investigations find that the diffusion of CRs in MHD turbulence is different from the standard diffusion model of CRs solely based on pitch-angle scattering \citep{Lazarian2021,Zhang2023,Hu2025,Kempski2025}. In addition to the pitch-angle scattering of CRs arising from the gyro-resonance with MHD turbulence, the magnetic mirroring effect leads to a new type of diffusion termed mirror diffusion, resulting in a L\'evy-flight-like propagation. In addition, using Monte-Carlo simulations, \citet{Liang2025} investigated the transport of cosmic rays in finite-size galaxies and found that cosmic rays experience anomalous diffusion on small scales and transition to normal diffusion on large scales.

As the alternative to the normal diffusion, the anomalous diffusion models of CRs have been proposed \citep{Uchaikin2013, Uchaikin2023, Zimbardo2017, Liang2025}. For instance, \citet{Lagutin2001} suggest that the ``Knee" (at about $3\times10^{15}$ eV) in the measured primary CR spectrum originates from the superdiffusion of CRs in the fractal ISM; \citet{Erlykin2013} propose that the superdiffusion can explain the CR spectrum being flatter in the Inner Galaxy than locally and in the Outer Galaxy; \citet{Wang2021} test recently the superdiffusion model in the ISM around the Geminga pulsar by investigating the surface brightness profile of TeV $\gamma$-ray halo associated with it. In addition, the superdiffusion is also applied to study the acceleration of energetic particles at shocks \citep{Effenberger2024, Aerdker2025}. Superdiffusive transport of energetic particles in the presence of magnetic turbulence can be ubiquitous in laboratory and astrophysical plasmas \citep[see, e.g.,][]{Zimbardo2015}. Indeed, there is evidence that energetic particles propagate superdiffusively in the heliosphere \citep{Perri2007, Perri2009} and in supernova remnants \citep{Perri2016} 

Since CRs measured locally almost do not deliver any information on their sources, we must resort to other probes for revealing their origin. Until present,  $\gamma$ rays, produced by the interactions of CRs with the ambient gas around their accelerators, are the most robust probe \citep{Drury1994}. Thus, it is crucial that we have a good understanding on the transport of CRs in the vicinity of their accelerators for explaining the observed $\gamma$ rays around them \citep{Aharonian1996}. In particular, the morphology of $\gamma$ rays provides the direct information on the injection history and spatial distribution of CRs around their accelerators.

%davancement of gamma-ray observations on CR distributions.
In this regard, thanks to the progress of detection technologies, $\gamma$-ray astronomy has achieved enough spatial resolution to investigate the morphology of $\gamma$-ray emissions. The spatial distribution of CRs can be derived by combining the  gas information and $\gamma$-ray morphology. With this method, H.E.S.S collaboration has derived the CR distribution in the central molecular zone (CMZ) in the Galactic Center region \citep{HESS2016}. The derived CR distribution obeys a $1/r$ distribution and is consistent with the continuous injection in the normal diffusion picture. Such $1/r$ distributions are also derived in several young massive star clusters (YMCs) \citep{Aharonian2019}. The future $\gamma$-ray instruments, especially the planned next generation imaging air cherenkov telescopes (IACTs) such as Cherenkov Telescope Array (CTA) \citep{CTA2019}, Astrofisica con Specchi a Tecnologia Replicante Italiana (ASTRI) \citep{ASTRI2022}, and Large Array of imaging atmospheric Cherenkov Telescope (LACT) \citep{LACT2026}, will further push the angular resolution of $\gamma$-ray astronomy down to 1 arcminute. Such high spatial resolution would  provide unique opportunities to test the CR propagation mechanisms beyond the normal diffusion in the vicinity of their accelerators.

In this paper, we calculate the distribution of CRs and the associated $\gamma$-ray emission near their accelerators within the framework of superdiffusion. The normal diffusion is included as a special case. For anomalous diffusion, the MSD scales nonlinearly with time, i.e., $\langle \Delta x^2 \rangle \propto t^\mu$, and superdiffusion corresponds to $\mu>1$, while subdiffusion corresponds to $\mu<1$. Anomalous diffusion can be described mathematically by the fractional calculus \citep{Metzler2000,Uchaikin2003}, and so also termed fractional diffusion. In the present work, we focus on investigating the superdiffusion of CRs in the vincinity of their sources. While subdiffusion \citep{Bouchaud1990} can also have interesting implications on the distribution of CRs near their sources and resulting $\gamma$-ray emission, we do not consider it in the present work. In addition, we also do not consider the ``Gaussianization", namely the transition to normal diffusion on large scales \citep{Liang2025}. Instead, we assume that the transport of CRs in the vicinity of their sources with a size of $\sim$100 pc is superdiffusive in nature. The paper is organized as follows: in Sec. \ref{sec:crdist}, we first discuss the Green's function of CR transport equation including energy losses, and next we compute the distribution of CR protons and electrons near their accelerators for impulsive and stationary injection, respectively, in Secs. \ref{sec:crimpulsive} and \ref{sec:crstationary}, at different ages for different diffusion coefficients; then in Sec. \ref{sec:gamma}, we compute the $\gamma$-ray emissions, which are produced by the interactions of CRs with ambient gas and radiation fields surrounding their accelerators, and discuss the temporal evolution and spatial distribution of $\gamma$ rays; finally, we summarize in Sec. \ref{sec:summary}.

%%%%%%%%%%%%%%%%%%%%%%%%%%%%%%%%%%%%%%%
%%%%% CR transport %%%%%%%%%%%%%%%%%%%%%%%%%%%
%%%%%%%%%%%%%%%%%%%%%%%%%%%%%%%%%%%%%%%
\section{The distributions of CR protons and electrons surrounding the accelerators} \label{sec:crdist}

The transport of CRs in the vicinity of their accelerators is described by the following equation \citep{Lagutin2001, Lagutin2023}

\begin{equation} \label{eq:transport}
   \frac{\partial N}{\partial t} + D(-\Delta)^{\alpha/2}N - \frac{\partial}{\partial p}[b(p) N] = Q,
\end{equation}

\noindent where $N(\mathbf{r}, p, t)$ is the number density of CRs per unit momentum, $D(p)$ is the spatial diffusion coefficient, $b(p) = -dp/dt$ is the momentum loss rate, and $Q(\mathbf{r}, p, t)$ is the source term. In the above equation, the fractional Laplacian operator $(-\Delta)^{\alpha/2}$ ($\Delta=\nabla^2$, $0 < \alpha \le 2$) is defined via Fourier transform \citep{Kwasnicki2017}

\begin{equation}
  \int \mathrm{e}^{\mathrm{i}\mathbf{k}\cdot\mathbf{r}}(-\Delta)^{\alpha/2}f d^3\mathbf{r} = k^\alpha \int f \mathrm{e}^{\mathrm{i}\mathbf{k}\cdot\mathbf{r}} d^3\mathbf{r},\quad k = \lvert\mathbf{k}\rvert,
\end{equation}

\noindent where $f(\mathbf{r})$ is an arbitrary function which has Fourier transform. When $\alpha=2$, we return to the normal Gaussian diffusion.

In Eq. \eqref{eq:transport}, the diffusion coefficient $D$, whose dimension is $\mathrm{L}^\alpha/\mathrm{T}$, is parameterized as \citep{Lagutin2003} 

\begin{equation} \label{eq:diffcoe}
  D(p) = D_0\beta \left(\frac{p}{1~\mathrm{GeV/c}}\right)^\delta,
\end{equation}

\noindent where $\beta$ is the ratio of CR speed to the light speed $c$.

We will assume that the gas around the accelerator of CRs is neutral. Given that the microphysics of CR transport is  incorporated into the diffusion coefficient, the phase of interstellar gas has no significant effects on the transport of CRs, since it only slightly influences energy losses of low-energy CRs with energies $\lesssim$ 1 GeV, while we are mainly interested in high-energy CRs.

\begin{figure}
    \centering
    \includegraphics[width=0.48\textwidth]{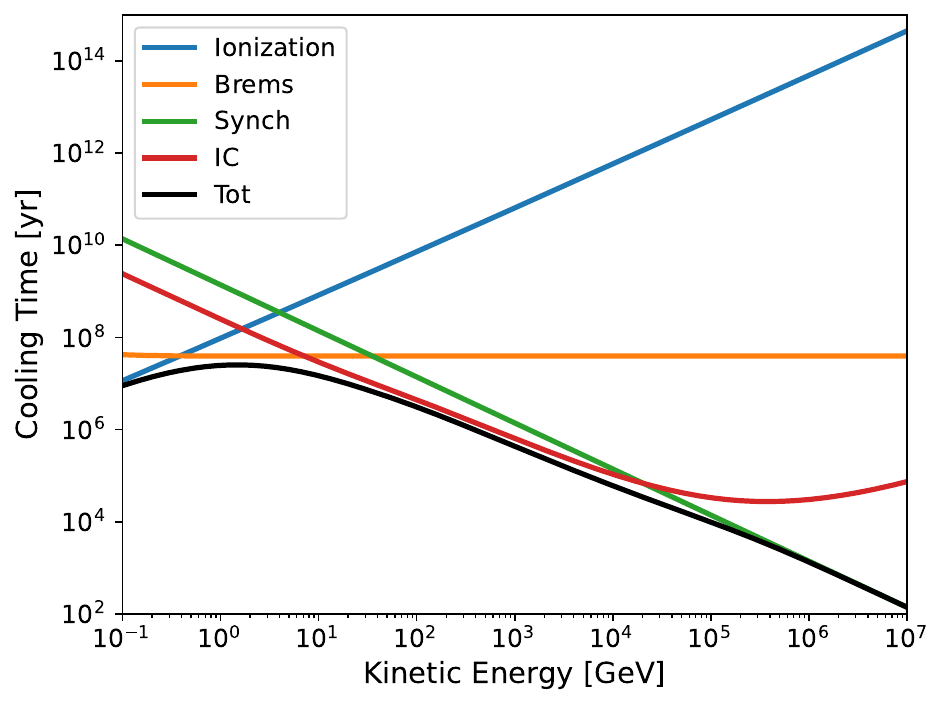}
    \includegraphics[width=0.48\textwidth]{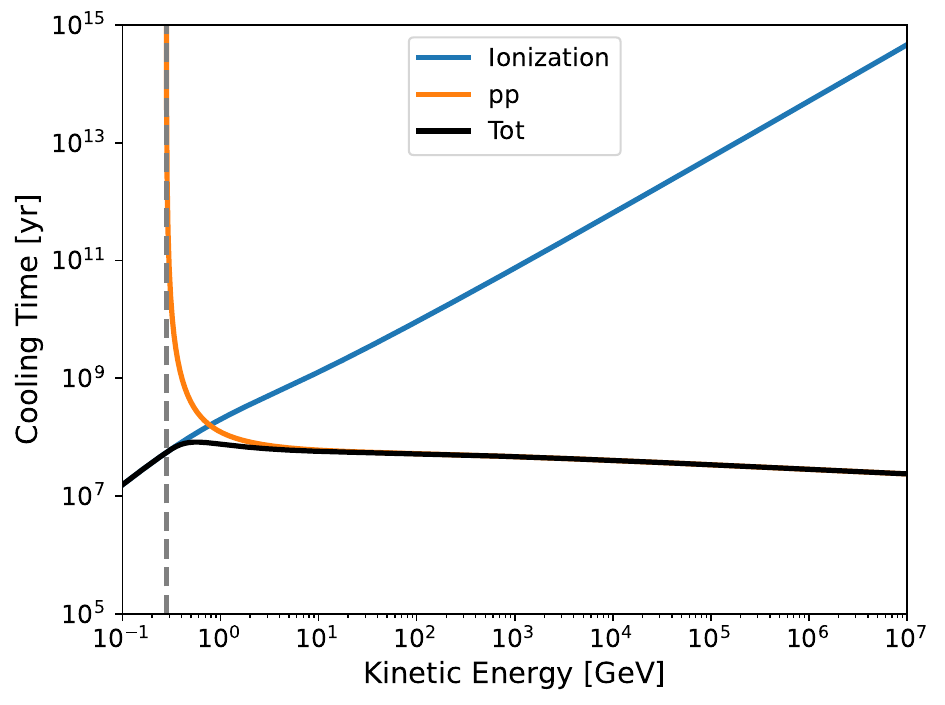}
    \caption{Cooling times of CR electrons (top) and protons (bottom) in a neutral gas. For computing the cooling times, we have assumed that the ambient gas hydrogen number density $n_\mathrm{H}=1.0~\mathrm{cm^{-3}}$, the ambient gas helium number density $n_\mathrm{He}=0.1n_\mathrm{H}$, and the magnetic field strength $B=3~\mathrm{\mu G}$. Moreover, for computing the IC scattering of electrons off the background soft photons due to CMB and ISRFs, we have assumed that the ISRFs consist of five diluted black-body radiation fields, whose temperatures are 23209.0, 6150.4, 3249.3, 313.3, and 33.1 K, and the corresponding energy densities are 0.12, 0.23, 0.37, 0.055, and 0.25 $\mathrm{eV/cm^3}$, respectively \citep{Delahaye2010, Fang2021}. In the right panel,  the vertical dashed line shows the kinetic energy  threshold, $E_\mathrm{k,th}=0.2797~\mathrm{GeV}$, for generating $\pi^0$-mesons from inelastic pp collisions.}
    \label{fig:cooltime}
\end{figure}

The dominant energy loss processes for CR electrons are due to ionization interactions with the ambient gas, bremsstrahlung, synchrotron radiation in the interstellar magnetic field, and inverse Compton (IC) scattering off background soft photons including cosmic microwave background (CMB) and interstellar radiation fields (ISRFs) \citep{Strong1998}. Klein-Nishina effect is taken into account in the IC scattering off a diluted black-body radiation field, and the corresponding energy loss has been parameterized \citep{Delahaye2010, Fang2021}. The top panel of Fig. \ref{fig:cooltime} shows the cooling times of electrons due to ionization (blue line), bremsstrahlung (orange line), synchrotron radiation (green line), and IC scattering (red line), and the total cooling time accounting for the all above processes (black line). For computing the cooling times, we have assumed that the ambient gas hydrogen number density $n_\mathrm{H} = 1.0~\mathrm{cm^{-3}}$, the ambient gas helium number density $n_\mathrm{He}=0.1n_\mathrm{H}$, and the magnetic field strength $B=3~\mathrm{\mu G}$. Moreover, the  ISRFs consist of five diluted black-body radiation fields, whose temperatures are 23209.0, 6150.4, 3249.3, 313.3, and 33.1 K, and the corresponding energy densities are 0.12, 0.23, 0.37, 0.055, and 0.25 $\mathrm{eV/cm^3}$, respectively \citep{Delahaye2010, Fang2021}.

The dominant energy loss processes for CR protons are due to ionization interactions and inelastic proton-proton (pp) collisions with the ambient gas \citep{Schlickeiser2002}. The lifetime of protons due to inelastic pp collisions is $\tau_{\mathrm{pp}}=1/[\kappa c\beta \sigma_{\mathrm{inel}} (n_\mathrm{H} + n_\mathrm{He}4^{0.79})]$, where the inelasticity $\kappa=0.5$ \citep{Aharonian2004}, $\sigma_\mathrm{inel}$ is the inelastic pp collision cross section \citep{Kafexhiu2014}, and the factor $4^{0.79}$ is a phenomenological correction to the inelastic cross section for helium nuclei \citep{Padovani2018}. The bottom panel of Fig. \ref{fig:cooltime} shows the cooling times of protons due to ionization (blue line) and inelastic pp collisions (orange line), and the total cooling time accounting for the above two processes (black line). For computing the above cooling times, we have assumed the same parameters as the top panel of Fig. \ref{fig:cooltime}. The vertical dashed line shows the kinetic energy threshold, $E_\mathrm{k,th}=0.2797~\mathrm{GeV}$, for generating $\pi^0$-mesons which decay promptly into two $\gamma$-ray photons from inelastic pp collisions.

We assume that around the accelerator the gas density and diffusion coefficient are homogeneous, and Eq. \eqref{eq:transport} is solved in the spherical symmetry with the accelerator in the center. The solution of Eq. \eqref{eq:transport} is 

\begin{equation} \label{eq:sol}
    N(\mathbf{r},p,t) = \int Q(\mathbf{r}', p', t') G(\mathbf{r}, p, t; \mathbf{r}', p', t') d^3\mathbf{r}'dp'dt',
\end{equation}

\noindent where the Green's function $G(\mathbf{r}, p, t; \mathbf{r}', p', t')$ satisfies 

\begin{equation}
    \frac{\partial G}{\partial t} + D(-\Delta)^{\alpha/2}G - \frac{\partial}{\partial p}[b(p) G] = \delta^{(3)}(\mathbf{r} - \mathbf{r}') \delta(p - p') \delta(t - t').
\end{equation}

\noindent Via applying Fourier transform for solving the above equation, we obtain Green's function as follows

\begin{equation} \label{eq:green}
\begin{aligned}
    G(\mathbf{r}, p, t; \mathbf{r}', p', t') &= \frac{1}{b(p)} \frac{g_3^{(\alpha)}[\lvert\mathbf{r} - \mathbf{r}'\rvert/\lambda^{1/\alpha}(p,p')]}{\lambda^{3/\alpha}(p, p')} \\ & \cdot \theta[\lambda(p, p')] \delta[t - t' - \tau(p, p')],
\end{aligned}
\end{equation}

\noindent where $\theta(x)$ is Heaviside step function, and we have introduced \citep{Syrovatskii1959}

\begin{equation}
  \tau(p, p') = \int_p^{p'}\frac{dp_1}{b(p_1)}, \quad \lambda(p, p') = \int_p^{p'}\frac{D(p_1)}{b(p_1)}dp_1.
\end{equation}

\noindent In Eq. \eqref{eq:green}, the function $g_3^{(\alpha)}(r)$ is defined by the following integral \citep{UchaikinZolotarev1999}

\begin{equation} \label{eq:g3alpha}
    g_3^{(\alpha)}(r) = \frac{1}{2\pi^2 r}\int_0^\infty \mathrm{e}^{-k^\alpha} \sin(kr) kdk.
\end{equation}

\noindent When $\alpha=1$,

\begin{equation} \label{eq:cauchy}
    g_3^{(1)}(r) = \frac{1}{\pi^2(r^2 + 1)^2},
\end{equation}

\noindent which is a Cauchy distribution. When $\alpha=2$, 

\begin{equation} \label{eq:gauss}
    g_3^{(2)}(r) = \frac{1}{(4\pi)^{3/2}}\mathrm{e}^{-r^2/4},
\end{equation}

\noindent which is a Gaussian distribution.

\begin{figure}
    \centering
    \includegraphics[width=0.5\textwidth]{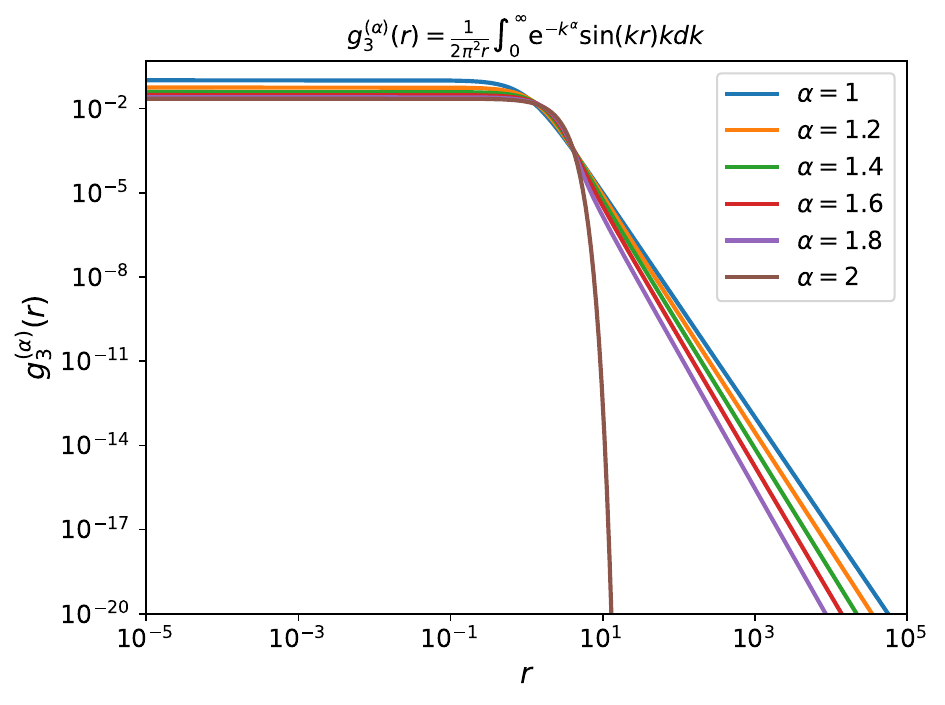}
    \caption{The function $g_3^{(\alpha)}(r)$ defined by Eq. \eqref{eq:g3alpha}.}
    \label{fig:g3alpha}
\end{figure}

Fig. \ref{fig:g3alpha} is a graph of the function $g_3^{(\alpha)}(r)$ for several different values of $\alpha$. When $r\ll 1$, we can see that $g_3^{(\alpha)}(r) \simeq \frac{1}{2\pi^2\alpha}\Gamma(3/\alpha)$, where $\Gamma(z)$ is the gamma function. However, when $r \gg 1$, $g_3^{(\alpha)}(r) \simeq \frac{1}{2\pi^2}\Gamma(\alpha+2)\sin(\alpha\pi/2)r^{-\alpha-3}$ for $\alpha < 2$ \citep{UchaikinZolotarev1999}, which displays a power-law tail and is significantly different from $g_3^{(2)}(r)$ given by Eq. \eqref{eq:gauss}.

%%%%%%%%%%%%%%%%%%%%%%%%%%%%%%%%%%%%
%%%%% impulsive injection %%%%%%%%%%%%%%%%%%%%%
%%%%%%%%%%%%%%%%%%%%%%%%%%%%%%%%%%%%
\subsection{Impulsive injection} \label{sec:crimpulsive}

We firstly consider that CRs are injected impulsively by an accelerator, for instance, by a supernova remnant (SNR) \citep{Aharonian1996}, into its surrounding ISM. For such impulsive injection, the source term of CRs is $Q(\mathbf{r},p,t) = q(p)\delta^{(3)}(\mathbf{r})\delta(t)$, where $q(p)$ is the momentum spectrum of injected CRs. Given the above source term, making use of Eqs. \eqref{eq:sol} and \eqref{eq:green}, the CR distribution is 

\begin{equation} \label{eq:impulsol}
  N(r, p, t) = \frac{b(p_0)}{b(p)} q(p_0) \frac{g_3^{(\alpha)}[r/\lambda^{1/\alpha}(p, p_0)]}{\lambda^{3/\alpha}(p, p_0)} \theta(p_b - p),
\end{equation}

\noindent where the initial momentum $p_0$ is given by $t = \tau(p, p_0)$, and the break momentum $p_b$ is given by $t = \tau(p_b, \infty)$.

Given that the cooling time of protons is generally much larger than the age of their accelerator and the diffusion time scale, we can see that $p_0 - p = b(p)t \ll p$, and $\lambda(p, p_0) = D(p)t$. Furthermore, since the cooling of protons at high energy is dominated by inelastic pp collisions, which is only weakly dependent on energy, we can see that $p_b = \infty$. As a consequence, we have \citep{Uchaikin2023}

\begin{equation} \label{eq:solnoloss-impulsive}
  N(r, p, t) = \frac{q(p)}{r_\mathrm{d}^3} g_3^{(\alpha)}(r/r_\mathrm{d}),
\end{equation}

\noindent where $r_\mathrm{d} = [D(p)t]^{1/\alpha}$. When $\alpha = 2$, $r_\mathrm{d} \propto t^{1/2}$, which results from Brownian motion and implies that CR particles move diffusively. When $\alpha < 2$,  on the other hand,  $r_\mathrm{d} \propto t^{1/\alpha}$, which results from L\'evy flights and implies that CR particles move superdiffusively \citep{Metzler2000,Chechkin2006}. Furthermore, when $\alpha < 2$, we have $N(r, p, t) \propto q(p)D(p)$ as $r \gg r_\mathrm{d}$.

CR electrons at high energy suffer severe energy losses due to synchrotron radiation and IC scattering. When considering the cooling due only to synchrotron radiation and IC scattering, and ignoring Klein-Nishina effect, $b(p) = b_0p^2$ \citep{Atoyan1995,Malyshev2009}, and thus we have $p_b =1/(b_0t)$. According to Eq. \eqref{eq:impulsol}, when $p>p_b$, $N(r, p, t) = 0$ implying that a sharp break forms in the energy spectrum of electrons due to cooling.

For simplicity, we will assume that the injection spectra of both CR protons ($i=$ p) and electrons ($i=$ e) are power-law, i.e., $q_{i}(p) \propto p^{-s}$. Moreover, we assume the same spectral index $s=2.2$ for both protons and electrons,  which is consistent with the inferences obtained from $\gamma$-ray observations to SNRs \citep{Caprioli2011} and from the local comic ray measurements \citep{Evoli2019}. The total kinetic energy injected into protons or electrons by the accelerator is $W_i = \int_{p_\mathrm{min}}^\infty q_i(p)E_{\mathrm{k},i}(p)dp$, where $p_\mathrm{min}=0.1~\mathrm{GeV/c}$ and $E_{\mathrm{k},i}(p)$ is the particle kinetic energy. For normalizing the injection spectra, we assume that $W_\mathrm{p} = 10^{50}~\mathrm{erg}$, which is about ten percent of the kinetic energy ($E_\mathrm{SN}\sim 10^{51}~\mathrm{erg}$) released by a typical supernova explosion \citep{Blasi2013, Batzofin2024}, and that the electron-to-proton ratio $K_\mathrm{ep}=W_\mathrm{e}/W_\mathrm{p}=0.001$, which is consistent with the canonical value got from observations to SNRs \citep{Morlino2012, Funk2013, Batzofin2024, Corso2023}.

The diffusion coefficient is crucial for determining the spatial and spectral distribution of CRs around their accelerators. Assuming that the diffusion of CRs in the Galactic ISM is described by the Brownian motion, namely that $\alpha=2$, the diffusion coefficient in the ISM is $D(1~\mathrm{GeV/c}) \sim 3\times10^{28}~\mathrm{cm^2/s}=0.1~\mathrm{pc^2/yr}$, based on the investigations on the propagation of CRs in the Galaxy and on the diffuse Galactic $\gamma$-ray emission \citep{Strong2007}. However, the diffusion exponent $\delta$ is not well constrained \citep{Strong2007, Silver2024}, due to a lack of detailed knowledge on the interstellar turbulence \citep{Grenier2015}. In the present work, we adopt $\delta=0.5$, in line with \citet{Aharonian1996}. On the other hand, the magnetic turbulence in the vicinity of CR accelerators can be significantly stronger than that in the diffuse ISM, and the nature of CR diffusion in the vicinity of accelerators can be anomalous, therefore, we adopt three different values for diffusion coefficient $D_0=$ 0.1, 0.01, and 0.001 $\mathrm{pc^\alpha/yr}$ for each value of $\alpha$, where $\alpha=$ 1, 1.2, 1.4, 1.6, 1.8 and 2 covering the possible parameter space.

\begin{figure*}
    \centering
    \includegraphics[width=0.99\textwidth]{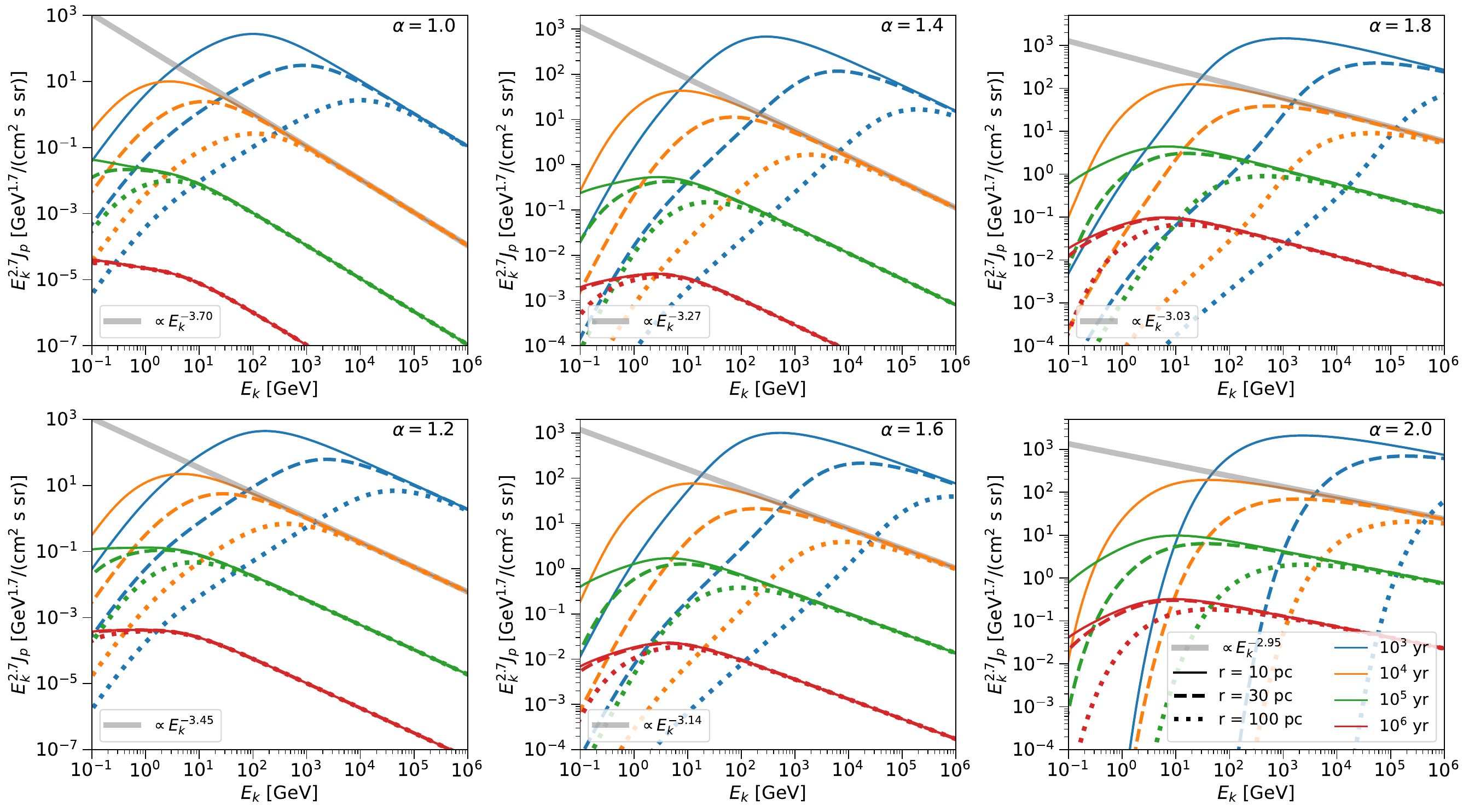}
    \caption{The energy spectra of CR protons at $t=10^3$ (blue lines), $10^4$ (orange lines), $10^5$ (green lines), and $10^6$ (red lines) years after the impulsive injection by an accelerator, when $D_0=0.001~\mathrm{pc^\alpha/yr}$ with $\alpha=$ 1, 1.2, 1.4, 1.6, 1.8, and 2 in each subplot from top to bottom and from left to right. The total kinetic energy injected into CR protons is $W_\mathrm{p}=10^{50}~\mathrm{erg}$, and the electron-to-proton ratio $K_\mathrm{ep}=0.001$. The ambient gas density, magnetic field strength, and ISRFs surrounding the accelerator are the same as those specified in discussing Fig. \ref{fig:cooltime}. In each subplot, the solid, dashed, and dotted lines show the spectra of protons at $r=$ 10, 30, and 100 pc, respectively; the gray line shows a power law with slope of $-(s+3\delta/\alpha)$, where the injection spectral index is $s=2.2$ and the diffusion exponent $\delta=0.5$.}
    \label{fig:pspec3-impulsive}
\end{figure*}

Fig. \ref{fig:pspec3-impulsive} displays the energy spectra, $J_\mathrm{p}(E_\mathrm{k,p})$, of CR protons at $t = 10^3$ (blue lines), $10^4$ (orange lines), $10^5$ (green lines), and $10^6$ (red lines) years after the accelerator injecting CRs impulsively, when $D_0=$ 0.001 $\mathrm{pc^\alpha/yr}$ with $\alpha=$ 1, 1.2, 1.4, 1.6, 1.8, and 2 for subplots from top to bottom and from left to right, respectively. In each subplot, the solid, dashed, and dotted lines are the spectra of protons at $r=$ 10, 30, and 100 pc, respectively. As expected, the flux of protons decreases with increasing $t$, for CRs are escaping from the accelerator. According to Eq. \eqref{eq:solnoloss-impulsive}, a prominent feature is that the spectra of protons are power-law with slope of $-(s+3\delta/\alpha)$ (gray lines) above a characteristic momentum, $p_\mathrm{p}^*$, which depends on the distance $r$, the diffusion coefficient $D_0$, and the superdiffusion parameter $\alpha$, and can be determined by the relation $r_\mathrm{d}(p_\mathrm{p}^*) = r$. The characteristic momentum $p_\mathrm{p}^*$ increases with increasing $r$ and decreasing $D_0$ for a given $\alpha$, and in particular, $p_\mathrm{p}^*$ increases with increasing $\alpha$ provided that both $r$ and $D_0$ are given. Below the characteristic momentum, the spectra are harder, and the flux of protons is lower at larger distance due to the momentum-dependent diffusion.\footnote{The above discussion holds for ultrarelativistic protons. Since the diffusion coefficient is dependent on $\beta$, and we have assumed that the injection spectrum is power-law in momentum, the spectral slope of protons with $E_\mathrm{k,p} \lesssim 1~\mathrm{GeV}$ is not necessarily $-(s+3\delta/\alpha)$ even though $r \ll r_\mathrm{d}$.} However, the distribution of protons depends on the distance more weakly when the superdiffusion parameter $\alpha$ is smaller, demonstrating the characteristic feature of superdiffusivity.

\begin{figure*}
    \centering
    \includegraphics[width=0.99\textwidth]{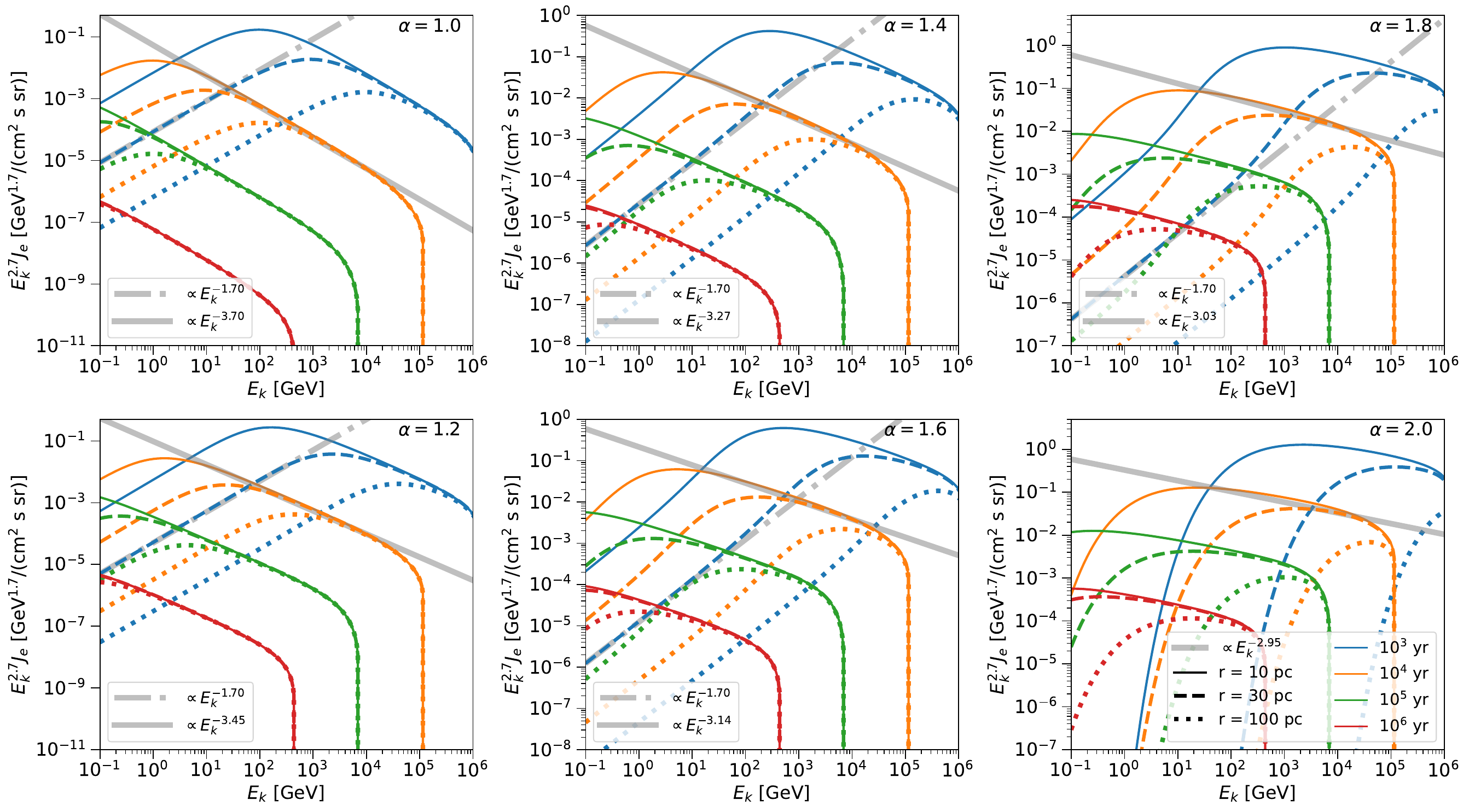}
    \caption{The energy spectra of CR electrons at $t=10^3$ (blue lines), $10^4$ (orange lines), $10^5$ (green lines), and $10^6$ (red lines) years after the impulsive injection by an accelerator, when $D_0=0.001~\mathrm{pc^\alpha/yr}$ with $\alpha=$ 1, 1.2, 1.4, 1.6, 1.8, and 2 in each subplot from top to bottom and from left to right. The total kinetic energy injected into CR protons is $W_\mathrm{p}=10^{50}~\mathrm{erg}$, and the electron-to-proton ratio $K_\mathrm{ep}=0.001$. The ambient gas density, magnetic field strength, and ISRFs surrounding the accelerator are the same as those specified in discussing Fig. \ref{fig:cooltime}. In each subplot, the solid, dashed, and dotted lines show the spectra of electrons at $r=$ 10, 30, and 100 pc, respectively; the gray solid line shows a power law with slope of $-(s+3\delta/\alpha)$, where the injection spectral index is $s=2.2$ and the diffusion exponent $\delta=0.5$. Moreover, except the bottom right subplot, the gray dotdashed line in each subplot shows a power law with slope of $-(s-\delta)$.}
    \label{fig:espec3-impulsive}
\end{figure*}

Fig. \ref{fig:espec3-impulsive} displays the energy spectra, $J_\mathrm{e}(E_\mathrm{k,e})$, of CR electrons at $t = 10^3$ (blue lines), $10^4$ (orange lines), $10^5$ (green lines), and $10^6$ (red lines) years after the accelerator injecting CRs impulsively, when $D_0=$ 0.001 $\mathrm{pc^\alpha/yr}$ with $\alpha=$ 1, 1.2, 1.4, 1.6, 1.8, and 2 for subplots from top to bottom and from left to right, respectively. In each subplot, the solid, dashed, and dotted lines are the spectra of electrons at $r=$ 10, 30, and 100 pc, respectively. As electrons suffer severe energy losses, we can see that the spectra are truncated above the cooling break $p_{b,\mathrm{e}}$, which decreases continuously with increasing $t$. When $cp_{\mathrm{e}}^{*} \ll E_\mathrm{k,e} \ll cp_{b,\mathrm{e}}$, similar to protons, the spectra of electrons are power-law above a characteristic momentum, $p_\mathrm{e}^*$, but below the cooling break, $p_{b,\mathrm{e}}$, and the spectral slope is $-(s+3\delta/\alpha)$ (gray solid lines). The characteristic momentum, $p_\mathrm{e}^*$, can be determined by the same relation $r_\mathrm{d}(p_\mathrm{e}^*)=r$ as that for protons, as long as $p_\mathrm{e}^* \ll p_{b,\mathrm{e}}$ and the energy losses are subdominant to diffusion. Furthermore, when $\alpha < 2$, the spectra of electrons are power-law with the slope of $-(s-\delta)$ as $E_\mathrm{k,e} \ll cp_\mathrm{e}^*$ (namely $r \gg r_\mathrm{d}$), significantly different from the spectral behavior for $\alpha=2$. Such a feature is the direct consequence of superdiffusivity, and in the subplots with $\alpha=$ 1, 1.2, 1.4, 1.6, and 1.8, the gray dotdashed lines show the power law with slope of $-(s-\delta)$.

\begin{figure*}
    \centering
    \includegraphics[width=0.99\textwidth]{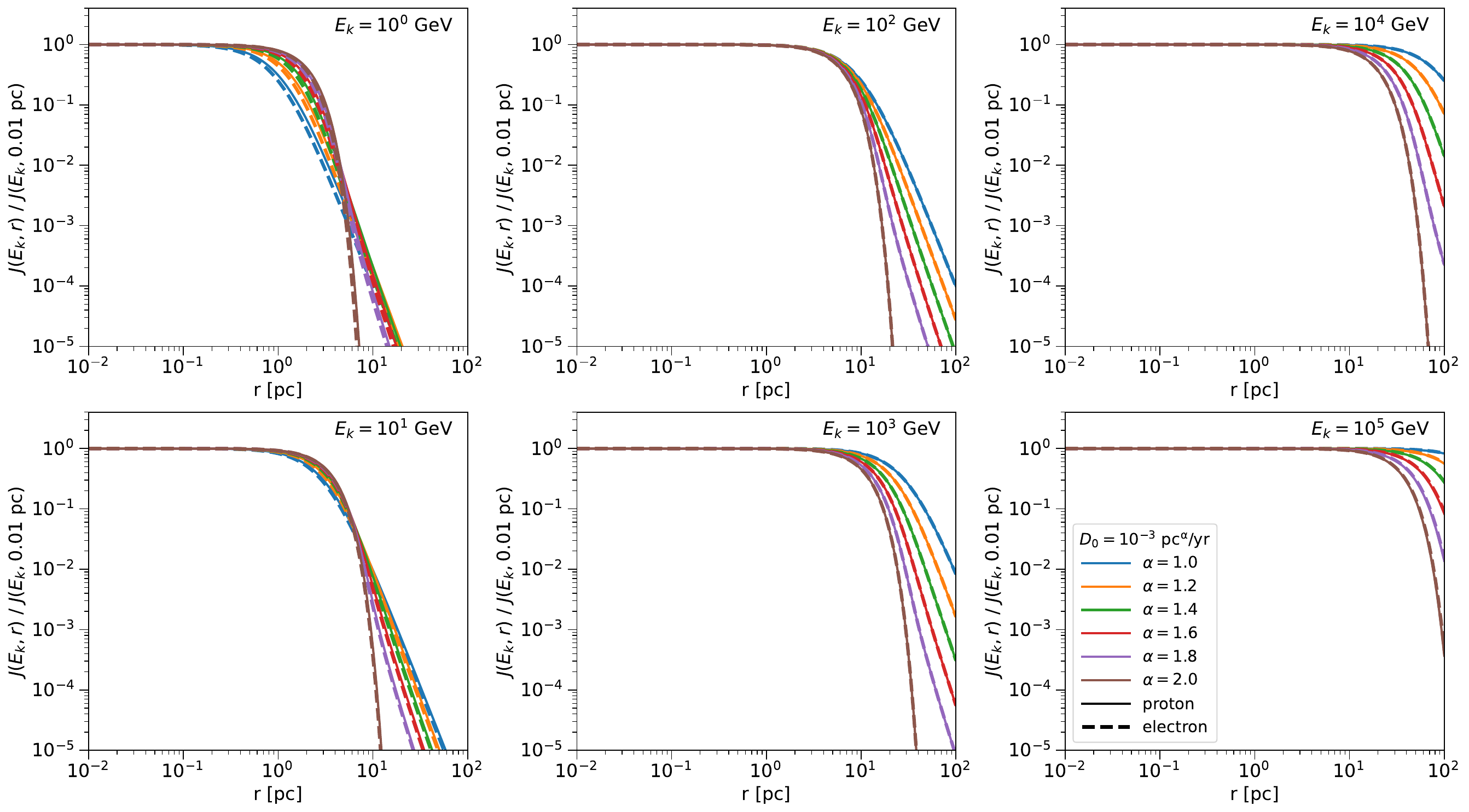}
    \caption{The radial profiles, which are normalized at $r=0.01$ pc, of CR protons (solid lines) and electrons (dashed lines) with $E_\mathrm{k}=1$ (top left panel), 10 (bottom left panel), $10^2$ (top middle panel), $10^3$ (bottom middle panel), $10^4$ (top right panel), and $10^5$ (bottom right panel) GeV at $t=10^3$ years after the impulsive injection by an accelerator, when $D_0=0.001~\mathrm{pc^\alpha/yr}$ with $\alpha=$ 1 (blue lines), 1.2 (orange lines), 1.4 (green lines), 1.6 (red lines), 1.8 (purple lines), and 2 (brown lines). The total kinetic energy injected into CR protons is $W_\mathrm{p}=10^{50}~\mathrm{erg}$, the electron-to-proton ratio $K_\mathrm{ep}=0.001$, the injection spectral index $s=2.2$, and the diffusion exponent $\delta=0.5$. The ambient gas density, magnetic field strength, and ISRFs surrounding the accelerator are the same as those specified in discussing Fig. \ref{fig:cooltime}.}
    \label{fig:crprof-impulsive}
\end{figure*}

Fig. \ref{fig:crprof-impulsive} displays the radial profiles of CR protons (solid lines) and electrons (dashed lines) at $t=10^3$ years after the accelerator injecting CRs impulsively, when $D_0=$ 0.001 $\mathrm{pc^\alpha/yr}$ with $\alpha=$ 1 (blue lines), 1.2 (orange lines), 1.4 (green lines), 1.6 (red lines), 1.8 (purple lines), and 2 (brown lines). The radial profile is defined by $J_i(E_{\mathrm{k},i},r)/J_i(E_{\mathrm{k},i},r_0)$ with $r_0=0.01$ pc, and we have chosen $E_{\mathrm{k},i}=$ 1, 10, $10^2$, $10^3$, $10^4$,and $10^5$ GeV for subplots from top to bottom and from left to right, respectively. The radial profile is determined by $g_3^{(\alpha)}(r/\lambda^{1/\alpha}(p,p_0))$, according to Eq. \eqref{eq:impulsol}, and it is constant as $r \ll \lambda^{1/\alpha}(p, p_0)$, while it is proportional to $r^{-\alpha - 3}$ for $\alpha < 2$ as $r \gg \lambda^{1/\alpha}(p, p_0)$. Given that $t=10^3$ years, the distribution of electrons is not affected significantly by the energy losses, and consequently electrons have nearly identical radial profiles with protons, while the radial profiles of the latter are in fact determined by $g_3^{(\alpha)}(r/r_\mathrm{d})$ in accordance with Eq. \eqref{eq:solnoloss-impulsive}. Furthermore, we can see that the radial profile is more extended for smaller $\alpha$ at a fixed $E_{\mathrm{k},i}$.

%%%%%%%%%%%%%%%%%%%%%%%%%%%%%%%%%%%%%%
%%%%%%%%%%%%%%%%%%%%%%%%%%%%%%%%%%%%%%
%\begin{figure}
%    \centering
%    \includegraphics[width=0.99\textwidth]{proton_spectra_impulsive_IK_D01e-1_s2.2.pdf}
%    \caption{The energy spectra of CR protons at $t=10^3$ (blue lines), $10^4$ (orange lines), $10^5$ (green lines), and $10^6$ (red lines) years after the impulsive injection by an accelerator, when $D_0=0.1~\mathrm{pc^\alpha/yr}$ with $\alpha=$ 1, 1.2, 1.4, 1.6, 1.8, and 2 in each subplot from top to bottom and from left to right. The total kinetic energy injected into CR protons is $W_\mathrm{p}=10^{50}~\mathrm{erg}$, and the electron-to-proton ratio $K_\mathrm{ep}=0.001$. The ambient gas density, magnetic field strength, and ISRFs surrounding the accelerator are the same as those specified in discussing Fig. \ref{fig:cooltime}. In each subplot, the solid, dashed, and dotted lines show the spectra of protons at $r=$ 10, 30, and 100 pc, respectively; the gray line shows a power law with slope of $-(s+3\delta/\alpha)$, where the injection spectral index is $s=2.2$ and the diffusion exponent $\delta=0.5$.}
%    \label{fig:pspec1-impulsive}
%\end{figure}

%\begin{figure}
%    \centering
%    \includegraphics[width=0.99\textwidth]{proton_spectra_impulsive_IK_D01e-2_s2.2.pdf}
%    \caption{Same as Fig. \ref{fig:pspec1-impulsive} but $D_0=0.01~\mathrm{pc^\alpha/yr}$.}
%    \label{fig:pspec2-impulsive}
%\end{figure}

%\begin{figure}
%    \centering
%    \includegraphics[width=0.99\textwidth]{electron_spectra_impulsive_IK_D01e-1_s2.2.pdf}
%    \caption{The energy spectra of CR electrons at $t=10^3$ (blue lines), $10^4$ (orange lines), $10^5$ (green lines), and $10^6$ (red lines) years after the impulsive injection by an accelerator, when $D_0=0.1~\mathrm{pc^\alpha/yr}$ with $\alpha=$ 1, 1.2, 1.4, 1.6, 1.8, and 2 in each subplot from top to bottom and from left to right. The total kinetic energy injected into CR protons is $W_\mathrm{p}=10^{50}~\mathrm{erg}$, and the electron-to-proton ratio $K_\mathrm{ep}=0.001$. The ambient gas density, magnetic field strength, and ISRFs surrounding the accelerator are the same as those specified in discussing Fig. \ref{fig:cooltime}. In each subplot, the solid, dashed, and dotted lines show the spectra of electrons at $r=$ 10, 30, and 100 pc, respectively; the gray solid line shows a power law with slope of $-(s+3\delta/\alpha)$, where the injection spectral index is $s=2.2$ and the diffusion exponent $\delta=0.5$. Moreover, except the bottom right subplot, the gray dotdashed line in each subplot shows a power law with slope of $-(s-\delta)$.}
%    \label{fig:espec1-impulsive}
%\end{figure}

%\begin{figure}
%    \centering
%    \includegraphics[width=0.99\textwidth]{electron_spectra_impulsive_IK_D01e-2_s2.2.pdf}
%    \caption{Same as Fig. \ref{fig:espec1-impulsive} but $D_0=0.01~\mathrm{pc^\alpha/yr}$.}
%    \label{fig:espec2-impulsive}
%\end{figure}

%%%%%%%%%%%%%%%%%%%%%%%%%%%%%%%%%%%%
%%%%%% stationary injection %%%%%%%%%%%%%%%%%%%%
%%%%%%%%%%%%%%%%%%%%%%%%%%%%%%%%%%%%
\subsection{Stationary injection} \label{sec:crstationary}

In this subsection, we investigate the cases where CRs are injected steadily by an accelerator, for instance, by the stellar wind of a massive star \citep{Seo2018}, or by a YMC \citep{Aharonian2019, Peron2024}, or by a microquasar \citep{Cao2025}, into its surrounding ISM after $t=0$. For such stationary injection, the source term of CRs is $Q(\mathbf{r}, p, t) = \dot{q}(p)\delta^{(3)}(\mathbf{r})\theta(t)$, where $\dot{q}(p)$ is the momentum spectra of injected CRs per unit time.  Given the above source term, making use of Eqs. \eqref{eq:sol} and \eqref{eq:green}, the CR distribution is

\begin{equation} \label{eq:statsol1}
   N(r, p, t) = \int_{t_0}^t \frac{b(p')}{b(p)} \frac{g_3^{(\alpha)}[r/\lambda^{1/\alpha}(p, p')]}{\lambda^{3/\alpha}(p, p')} \dot{q}(p') dt', 
\end{equation}

\noindent where the momentum $p'$ is determined by the relation $t - t' = \tau(p, p')$, and $t_0 = \mathrm{max}\{0, t - \tau(p, \infty)\}$. The above solution can be expressed more conveniently by the following

\begin{equation} \label{eq:statsol2}
  N(r, p, t) = \frac{1}{b(p)} \int_p^{p_0} \frac{g_3^{(\alpha)}[r/\lambda^{1/\alpha}(p,p')]}{\lambda^{3/\alpha}(p, p')} \dot{q}(p') dp',
\end{equation}

\noindent where the upper momentum $p_0$ is determined by the relation $t=\tau(p,p_0)$ when $p < p_b$, and $p_0=\infty$ otherwise. The break momentum $p_b$ is defined by $t=\tau(p_b,\infty)$.

When the cooling time of CRs is larger than the age of their accelerator and the diffusion time scale, Eq. \eqref{eq:statsol1} is reduced to

\begin{equation} \label{eq:statsolnoloss1}
  N(r, p, t) = \int_{0}^{t} \frac{g_3^{(\alpha)}[r/[D(p)(t-t')]^{1/\alpha}]}{[D(p)(t-t')]^{3/\alpha}} \dot{q}(p) dt'.
\end{equation}

\noindent The above equation can be also expressed as

\begin{equation} \label{eq:statsolnoloss2}
  N(r, p, t) = \frac{\dot{q}(p)}{D(p)}\frac{\alpha}{r^{3-\alpha}} \int_{r/r_\mathrm{d}}^{\infty}\xi^{2-\alpha}g_3^{(\alpha)}(\xi)d\xi,
\end{equation}

\noindent where $r_\mathrm{d} = [D(p)t]^{1/\alpha}$. In particular,  when $\alpha=1$, we have

\begin{equation}
  N(r, p, t) = \frac{\dot{q}(p)}{D(p)}\frac{1}{2\pi^2r^2 [(r/r_\mathrm{d})^2+1]},
\end{equation}

\noindent and when $\alpha=2$, we have \citep{Atoyan1995, Aharonian1996}

\begin{equation}
  N(r, p, t) = \frac{\dot{q}(p)}{D(p)}\frac{\mathrm{erfc}(r/2r_\mathrm{d})}{4\pi r},
\end{equation}

\noindent where $\mathrm{erfc}(x)$ is the  complementary error function (when $x\gg1$, $\mathrm{erfc}(x)\simeq \frac{1}{x\sqrt{\pi}}\mathrm{e}^{-x^2}$). When $\alpha < 2$, we have $N(r, p, t) \propto \dot{q}(p)D(p)$ as $r \gg r_\mathrm{d}$. It is interesting to see that $N(r, p, t) \propto r^{\alpha - 3}$, when $r \ll r_\mathrm{d}$. The radial dependence of CR distribution on the superdiffusion parameter $\alpha$ should have significant impacts on the morphology of $\gamma$-ray emission, which is generated by the interactions of CRs with the surrounding ISM and ISRFs of their accelerator.

Similar to the impulsive injection, we will assume that the injection spectra of CR protons ($i=$ p) and electrons ($i=$ e) are power-law, $\dot{q}_i(p)\propto p^{-s}$, with the spectral index $s=2.2$. The total kinetic power injected into protons or electrons by the accelerator is $\dot{W}_i = \int_{p_\mathrm{min}}^\infty \dot{q}_i(p)E_{\mathrm{k},i}(p)dp$, where $p_{\mathrm{min}}=0.1~\mathrm{GeV/c}$. For normalizing the injection spectra, we assume that $\dot{W}_\mathrm{p}=10^{37}~\mathrm{erg/s}$, and the electron-to-proton ratio $K_\mathrm{ep}=\dot{W}_\mathrm{e}/\dot{W}_\mathrm{p}=0.001$. The typical kinetic power of YMCs is about $10^{38}~\mathrm{erg/s}$, and the kinetic power of microquasars can be as high as $10^{39}~\mathrm{erg/s}$ \citep{Wang2025}, thus the above normalization is plausible.

\begin{figure*}
    \centering
    \includegraphics[width=0.99\textwidth]{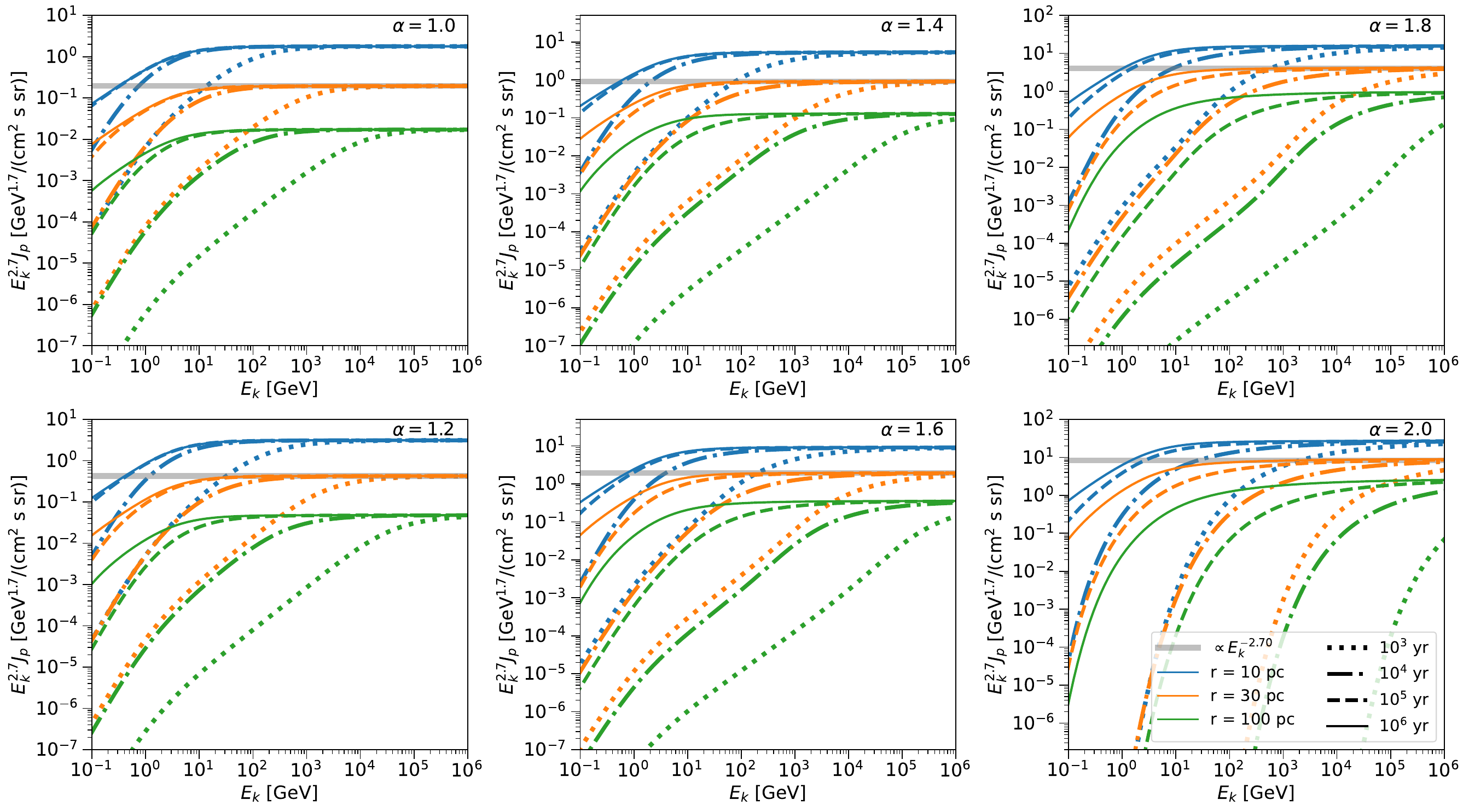}
    \caption{The energy spectra of CR protons at $t=10^3$ (dotted lines), $10^4$ (dotdashed lines), $10^5$ (dashed lines), and $10^6$ (solid lines) years after the stationary injection starting from $t=0$ by an accelerator, when $D_0=0.001~\mathrm{pc^\alpha/yr}$ with $\alpha=$ 1, 1.2, 1.4, 1.6, 1.8, and 2 in each subplot from top to bottom and from left to right. The total kinetic power injected into CR protons is $\dot{W}_\mathrm{p}=10^{37}~\mathrm{erg/s}$, and the electron-to-proton ratio $K_\mathrm{ep}=0.001$. The ambient gas density, magnetic field strength, and ISRFs surrounding the accelerator are the same as those specified in discussing Fig. \ref{fig:cooltime}. In each subplot, the blue, orange, and green lines show the spectra of protons at $r=$ 10, 30, and 100 pc, respectively; the gray line shows a power law with slope of $-(s+\delta)$, where the injection spectral index is $s=2.2$ and the diffusion exponent $\delta=0.5$.}
    \label{fig:pspec3-stationary}
\end{figure*}

Fig. \ref{fig:pspec3-stationary} displays the energy spectra, $J_\mathrm{p}(E_\mathrm{k,p})$, of CR protons at $t=10^3$ (dotted lines), $10^4$ (dotdashed lines), $10^5$ (dashed lines), and $10^6$ (solid lines) years after the accelerator injecting CRs steadily, when $D_0=$ 0.001 $\mathrm{pc^\alpha/yr}$ with $\alpha=$ 1, 1.2, 1.4, 1.6, 1.8, and 2 for subplots from top to bottom and from left to right, respectively. In each subplot, the blue, orange, and green lines show the spectra of protons at $r=$ 10, 30, and 100 pc, respectively. According to Eq. \ref{eq:statsolnoloss2}, the spectra are power-law with slope of $-(s+\delta)$ (gray lines) above a characteristic momenum, $p_\mathrm{p}^*$, which can be determined by the relation $r_\mathrm{d}(p_\mathrm{p}^*)=r$ similar to the impulsive injection. Moreover, at a given distance $r$, the spectra will be time-independent once $r \ll r_\mathrm{d}$, demonstrating the interplay of replenishment of CRs due to continuous injection and escaping due to diffusion. When $\alpha$ is smaller, we can see that the spectra approach more quickly the time-independent ones.

\begin{figure*}
    \centering
    \includegraphics[width=0.99\textwidth]{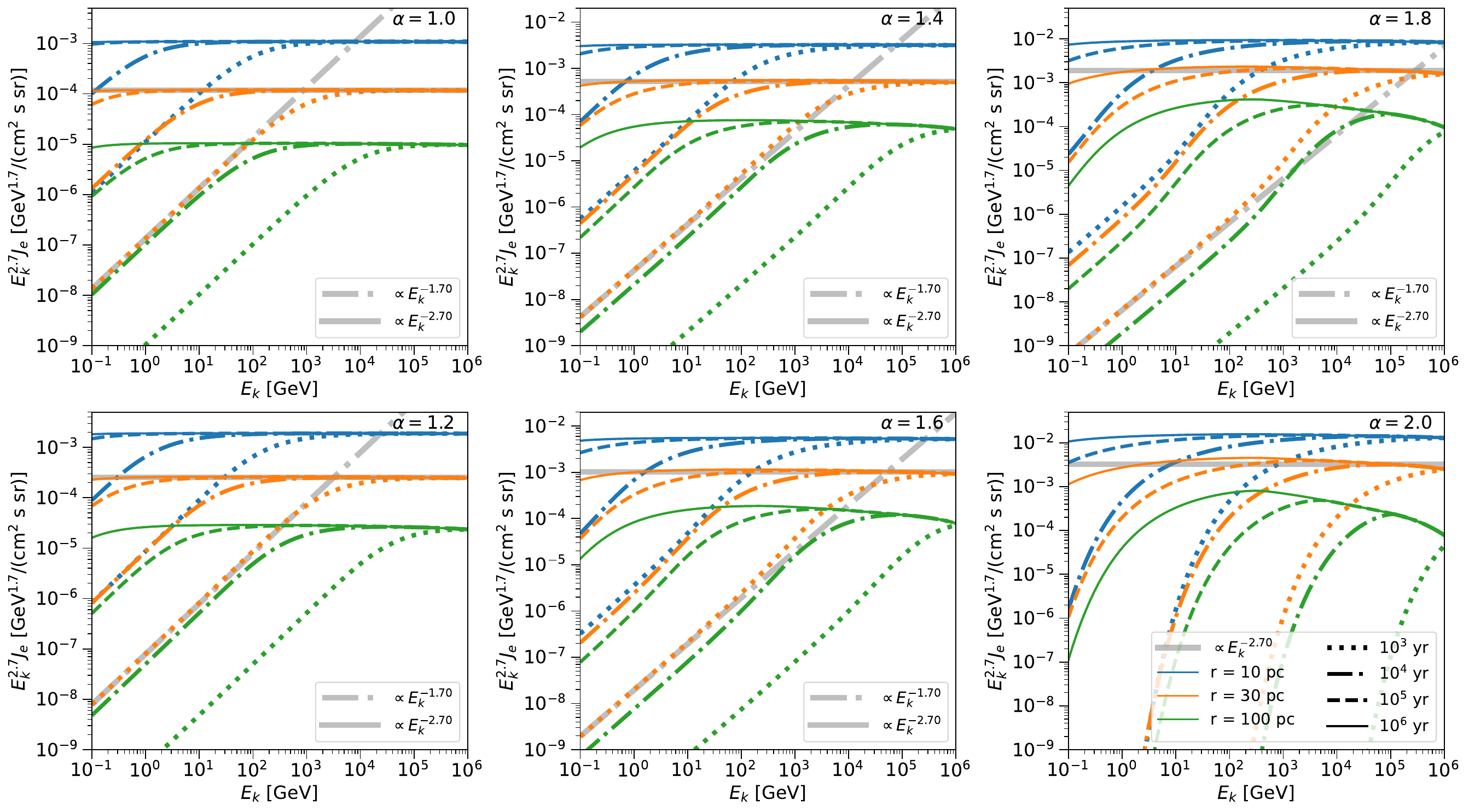}
    \caption{The energy spectra of CR electrons at $t=10^3$ (dotted lines), $10^4$ (dotdashed lines), $10^5$ (dashed lines), and $10^6$ (solid lines) years after the stationary injection starting from $t=0$ by an accelerator, when $D_0=0.001~\mathrm{pc^\alpha/yr}$ with $\alpha=$ 1, 1.2, 1.4, 1.6, 1.8, and 2 in each subplot from top to bottom and from left to right. The total kinetic power injected into CR protons is $\dot{W}_\mathrm{p}=10^{37}~\mathrm{erg/s}$, and the electron-to-proton ratio $K_\mathrm{ep}=0.001$. The ambient gas density, magnetic field strength, and ISRFs surrounding the accelerator are the same as those specified in discussing Fig. \ref{fig:cooltime}. In each subplot, the blue, orange, and green lines show the spectra of electrons at $r=$ 10, 30, and 100 pc, respectively; the gray solid line shows a power law with slope of $-(s+\delta)$, where the injection spectral index is $s=2.2$ and the diffusion exponent $\delta=0.5$. Moreover, except the bottom right subplot, the gray dotdashed line in each subplot shows a power law with slope of $-(s-\delta)$.}
    \label{fig:espec3-stationary}
\end{figure*}

Fig. \ref{fig:espec3-stationary} displays the energy spectra, $J_\mathrm{e}(E_\mathrm{k,e})$, of CR electrons at $t=10^3$ (dotted lines), $10^4$ (dotdashed lines), $10^5$ (dashed lines), and $10^6$ (solid lines) years after the accelerator injecting CRs steadily, when $D_0=$ 0.001 $\mathrm{pc^\alpha/yr}$ with $\alpha=$ 1, 1.2, 1.4, 1.6, 1.8, and 2 for subplots from top to bottom and from left to right, respectively. In each subplot, the blue, orange, and green lines show the spectra of electrons at $r=$ 10, 30, and 100 pc, respectively; the gray solid line shows a power law with slope of $-(s+\delta)$. When $D_0$ is smaller, as shown in Fig. \ref{fig:espec3-stationary}, the spectral slope is different from $-(s+\delta)$ at high energies, and the difference becomes larger for larger $r$, since the energy losses play a more important role. Nevertheless, it is interesting to note that the difference is smaller, when $\alpha$ is smaller. On the other hand, when $D_0$ is larger (not shown here), namely when diffusion becomes more important, the spectral slope will tend to $-(s+\delta)$ above the characteristic momentum, $p_\mathrm{e}^*$, which is determined by $r_\mathrm{d}(p_\mathrm{e}^*)=r$. Furthermore, when $\alpha < 2$ and  as $E_\mathrm{k,e} \ll cp_\mathrm{e}^*$ (namely $r \gg r_\mathrm{d}$), since $g_3^{(\alpha)}(\xi) \propto \xi^{-(\alpha+3)}$ and accordingly the integration in Eq. \eqref{eq:statsolnoloss2} is proportional to $(r_\mathrm{d}/r)^{2\alpha}$, the spectra are power-law with slope of $-(s-\delta)$ (gray dotdashed lines).

\begin{figure*}
    \centering
    \includegraphics[width=0.99\textwidth]{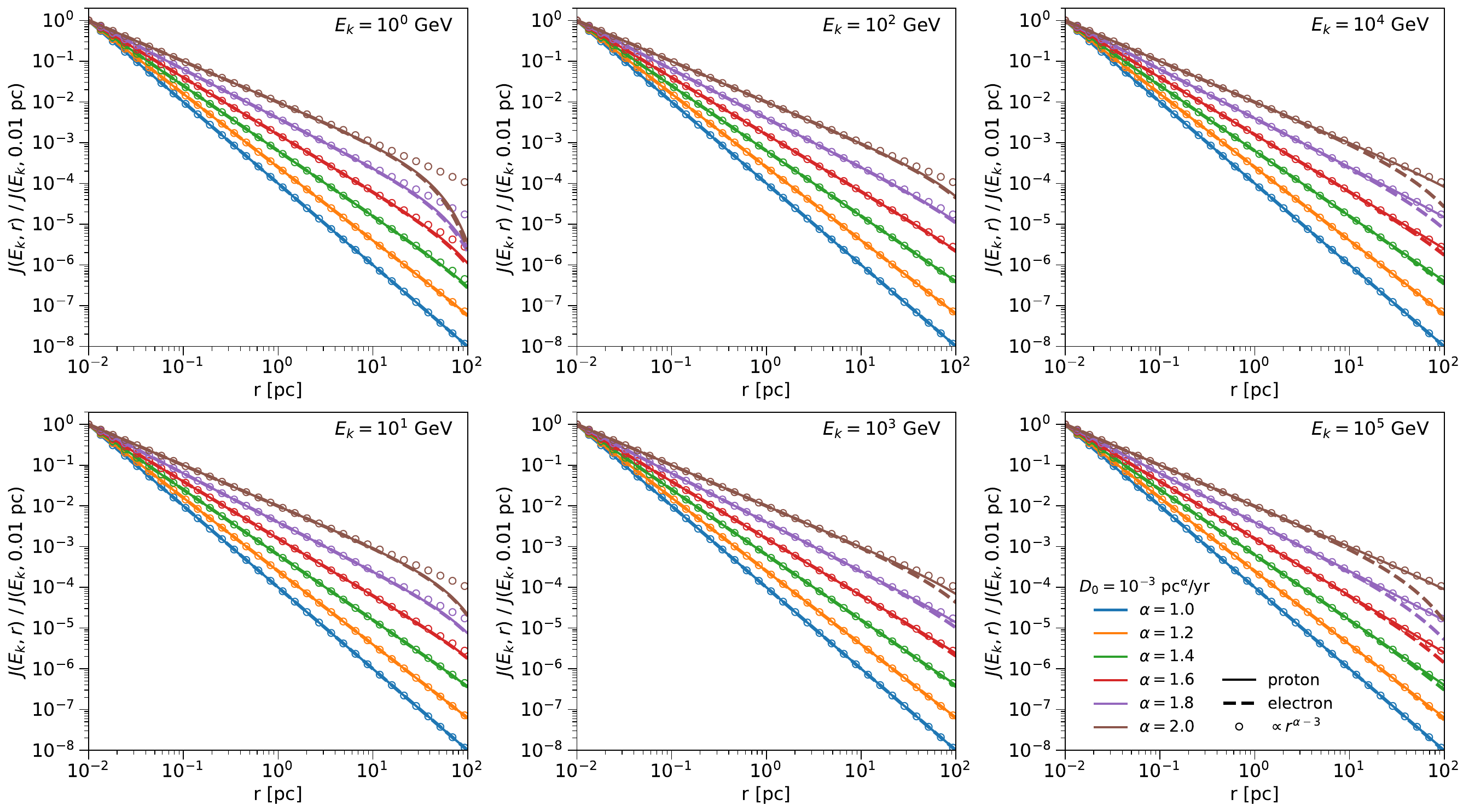}
    \caption{The radial profiles, which are normalized at $r=0.01$ pc, of CR protons (solid lines) and electrons (dashed lines) with $E_\mathrm{k}=1$ (top left panel), 10 (bottom left panel), $10^2$ (top middle panel), $10^3$ (bottom middle panel), $10^4$ (top right panel), and $10^5$ (bottom right panel) GeV at $t=10^6$ years after the stationary injection starting from $t=0$ by an accelerator, when $D_0=0.001~\mathrm{pc^\alpha/yr}$ with $\alpha=$ 1 (blue lines), 1.2 (orange lines), 1.4 (green lines), 1.6 (red lines), 1.8 (purple lines), and 2 (brown lines). In each subplot, the open circles show the radial profiles proportional to $r^{\alpha - 3}$. The total kinetic power injected into CR protons is $\dot{W}_\mathrm{p}=10^{37}~\mathrm{erg/s}$, the electron-to-proton ratio $K_\mathrm{ep}=0.001$, the injection spectral index $s=2.2$, and the diffusion exponent $\delta=0.5$. The ambient gas density, magnetic field strength, and ISRFs surrounding the accelerator are the same as those specified in discussing Fig. \ref{fig:cooltime}.}
    \label{fig:crprof-stationary}
\end{figure*}

Fig. \ref{fig:crprof-stationary} displays the radial profiles of CR protons (solid lines) and electrons (dashed lines) at $t=10^6$ years after the accelerator injecting CRs steadily, when $D_0=$ 0.001 $\mathrm{pc^\alpha/yr}$ with $\alpha=$ 1 (blue lines), 1.2 (orange lines), 1.4 (green lines), 1.6 (red lines), 1.8 (purple lines), and 2 (brown lines). Similar to the impulsive injection, the radial profile is defined by $J_i(E_{\mathrm{k},i},r)/J_i(E_{\mathrm{k},i},r_0)$ with $r_0=0.01$ pc, and we have chosen $E_{\mathrm{k},i}=1$, 10, $10^2$, $10^3$, $10^4$, and $10^5$ GeV for subplots from top to bottom and from left to right, respectively. For protons, their radial profiles are determined in accordance with Eq. \eqref{eq:statsolnoloss2}, and when $r\ll r_\mathrm{d}$, namely when $E_\mathrm{k,p}$ is high enough, their radial profile will be proportional to $r^{\alpha - 3}$, as shown via open circles. On the other hand, due to the severe energy losses, the radial profiles of electrons do not follow $r^{\alpha - 3}$ generally at high energies, especially when $\alpha$ is larger.

%%%%%%%%%%%%%%%%%%%%%%%%%%%%%%%%%%%%%%
%%%%%%%%%%%%%%%%%%%%%%%%%%%%%%%%%%%%%%
%\begin{figure}
%    \centering
%    \includegraphics[width=0.99\textwidth]{proton_spectra_stationary_IK_D01e-1_s2.2.pdf}
%    \caption{The energy spectra of CR protons at $t=10^3$ (dotted lines), $10^4$ (dotdashed lines), $10^5$ (dashed lines), and $10^6$ (solid lines) years after the stationary injection starting from $t=0$ by an accelerator, when $D_0=0.1~\mathrm{pc^\alpha/yr}$ with $\alpha=$ 1, 1.2, 1.4, 1.6, 1.8, and 2 in each subplot from top to bottom and from left to right. The total kinetic power injected into CR protons is $\dot{W}_\mathrm{p}=10^{37}~\mathrm{erg/s}$, and the electron-to-proton ratio $K_\mathrm{ep}=0.001$. The ambient gas density, magnetic field strength, and ISRFs surrounding the accelerator are the same as those specified in discussing Fig. \ref{fig:cooltime}. In each subplot, the blue, orange, and green lines show the spectra of protons at $r=$ 10, 30, and 100 pc, respectively; the gray line shows a power law with slope of $-(s+\delta)$, where the injection spectral index is $s=2.2$ and the diffusion exponent $\delta=0.5$.}
%    \label{fig:pspec1-stationary}
%\end{figure}

%\begin{figure}
%    \centering
%    \includegraphics[width=0.99\textwidth]{proton_spectra_stationary_IK_D01e-2_s2.2.pdf}
%    \caption{Same as Fig. \ref{fig:pspec1-stationary} but $D_0=0.01~\mathrm{pc^\alpha/yr}$.}
%    \label{fig:pspec2-stationary}
%\end{figure}

%\begin{figure}
%    \centering
%    \includegraphics[width=0.99\textwidth]{electron_spectra_stationary_IK_D01e-1_s2.2.pdf}
%    \caption{The energy spectra of CR electrons at $t=10^3$ (dotted lines), $10^4$ (dotdashed lines), $10^5$ (dashed lines), and $10^6$ (solid lines) years after the stationary injection starting from $t=0$ by an accelerator, when $D_0=0.1~\mathrm{pc^\alpha/yr}$ with $\alpha=$ 1, 1.2, 1.4, 1.6, 1.8, and 2 in each subplot from top to bottom and from left to right. The total kinetic power injected into CR protons is $\dot{W}_\mathrm{p}=10^{37}~\mathrm{erg/s}$, and the electron-to-proton ratio $K_\mathrm{ep}=0.001$. The ambient gas density, magnetic field strength, and ISRFs surrounding the accelerator are the same as those specified in discussing Fig. \ref{fig:cooltime}. In each subplot, the blue, orange, and green lines show the spectra of electrons at $r=$ 10, 30, and 100 pc, respectively; the gray solid line shows a power law with slope of $-(s+\delta)$, where the injection spectral index is $s=2.2$ and the diffusion exponent $\delta=0.5$. Moreover, except the bottom right subplot, the gray dotdashed line in each subplot shows a power law with slope of $-(s-\delta)$.}
%    \label{fig:espec1-stationary}
%\end{figure}

%\begin{figure}
%    \centering
%    \includegraphics[width=0.99\textwidth]{electron_spectra_stationary_IK_D01e-2_s2.2.pdf}
%    \caption{Same as Fig. \ref{fig:espec1-stationary} but $D_0=0.01~\mathrm{pc^\alpha/yr}$.}
%    \label{fig:espec2-stationary}
%\end{figure}

%%%%%%%%%%%%%%%%%%%%%%%%%%%%%%%%%%%%%%%
%%%%% gamma-ray %%%%%%%%%%%%%%%%%%%%%%%%%%%%
%%%%%%%%%%%%%%%%%%%%%%%%%%%%%%%%%%%%%%%
\section{$\gamma$-ray emission surrounding the accelerators} \label{sec:gamma}

Once we know the distributions of CR protons and electrons surrounding the accelerators, we can compute the emissivities of $\gamma$ rays resulting from the decay of $\pi^0$-mesons, which are produced by inelastic collisions of CR protons with ambient gas, from bremsstrahlung of CR electrons, and from IC scattering of CR electrons off background soft photons including CMB and ISRFs. For computing the $\gamma$-ray emissivities, we make use of the package \textsc{Naima} \citep{naima}. In \textsc{Naima}, the $\pi^0$-decay $\gamma$-ray emissivity is computed by implementing the parameterization of $\gamma$-ray production proposed by \citet{Kafexhiu2014}, the bremsstrahlung $\gamma$-ray emissivity is computed by using the expressions given in \citet{Baring1999}, and the IC $\gamma$-ray emissivity is computed by implementing the parameterization of \citet{Khangulyan2014}. Furthermore, when computing the $\pi^0$-decay emissivity, we assume a nuclear enhancement factor $\varepsilon_\mathrm{M}=1.8$ accounting for the contribution to the $\gamma$-ray production of nuclei heavier than hydrogen in both CRs and ambient gas \citep{Mori2009, Kachelriess2014}.

The $\gamma$-ray flux and intensity, respectively, are

\begin{equation}
  F_\gamma(E_\gamma) = \frac{1}{d^2} \int_0^R q_\gamma(E_\gamma, r) 4\pi r^2 dr,
\end{equation}

\noindent and

\begin{equation}
  I_\gamma(E_\gamma, \vartheta) = 2\int_0^{\sqrt{R^2 - d^2\sin^2\vartheta}}q_\gamma(E_\gamma, \sqrt{l^2 + d^2\sin^2\vartheta})dl,
\end{equation}

\noindent where $q_\gamma(E_\gamma, r)$ is the emissivity (in units of $\mathrm{cm^{-3}~s^{-1}~sr^{-1}~GeV^{-1}}$), $d$ is the distance to the accelerator, $R$ is the dimension of $\gamma$-ray emission volume, and $\vartheta$ is the angular separation from the accelerator. We will assume that $d=$ 1 kpc, and $R=$ 100 pc. Furthermore, we will concentrate on the $\gamma$ rays with energies above 100 MeV. In the following, we first discuss the cases for impulsive injection, and then for stationary injection.

\begin{figure*}
    \centering
    \includegraphics[width=0.99\textwidth]{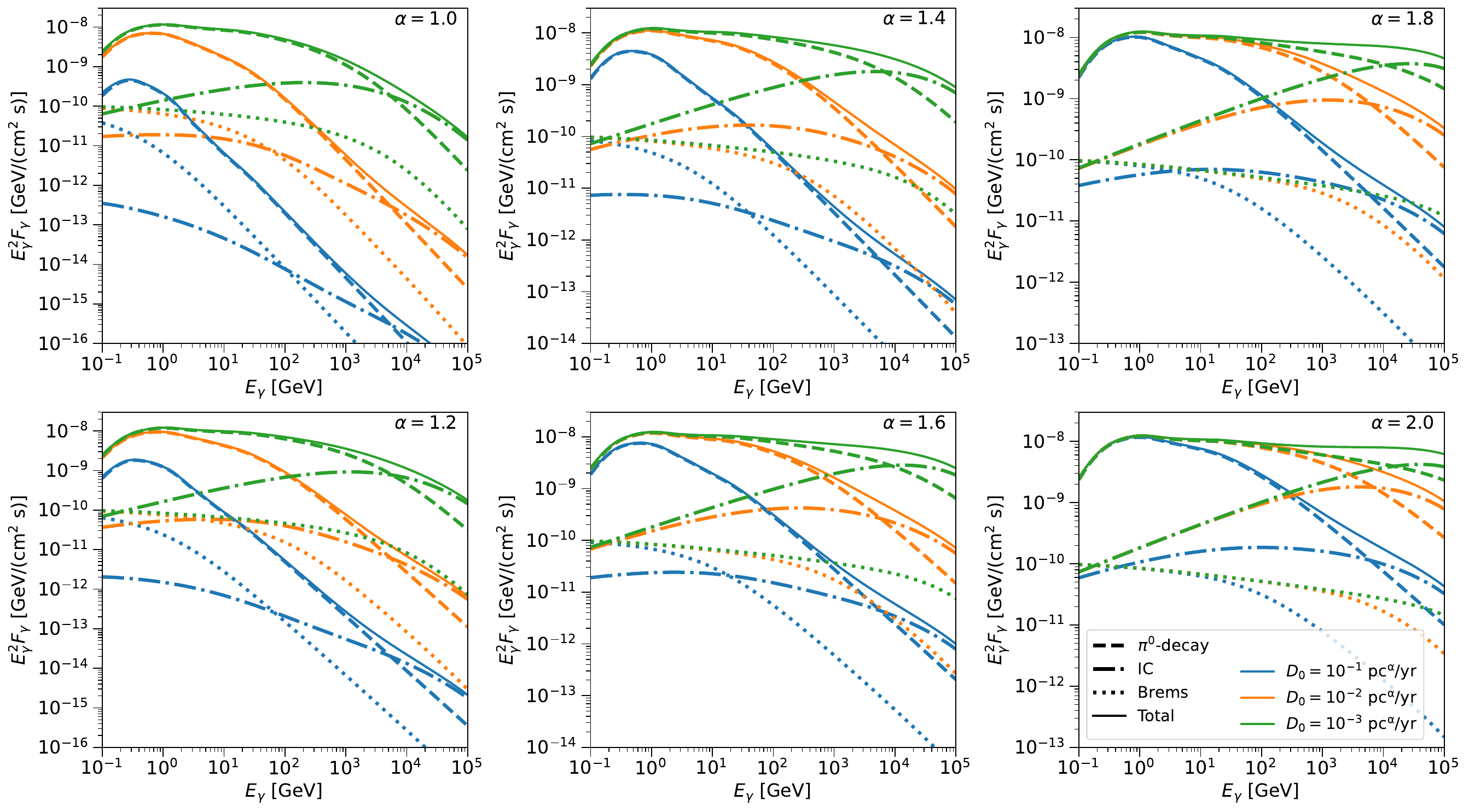}
    \caption{The $\gamma$-ray fluxes at $t=10^3$ years after the impulsive injection by an accelerator, when $D_0=$ 0.1 (blue lines), 0.01 (orange lines), and 0.001 (green lines) $\mathrm{pc^\alpha/yr}$ with $\alpha=$ 1, 1.2, 1.4, 1.6, 1.8, and 2 in each subplot from top to bottom and from left to right. The distance to the accelerator is $d=$ 1 kpc, and the dimension of $\gamma$-ray emission volume is $R=$ 100 pc. The total kinetic energy injected into CR protons is $W_\mathrm{p}=10^{50}~\mathrm{erg}$, the electron-to-proton ratio $K_\mathrm{ep} = 0.001$, the injection spectral index $s=2.2$ for both CR protons and electrons, and the diffusion exponent $\delta=0.5$. The ambient gas density, magnetic field strength, and ISRFs surrounding the accelerator are the same as those specified in discussing Fig. \ref{fig:cooltime}. In each subplot, the dashed, dotdashed, and dotted lines show the fluxes of $\gamma$ rays generated from the decay of $\pi^0$-mesons, IC, and bremsstrahlung, respectively, while the solid lines show the total fluxes.}
    \label{fig:gspec1-impulsive}
\end{figure*}

Fig. \ref{fig:gspec1-impulsive} shows the fluxes of $\gamma$ rays produced by the interactions of CRs with ambient gas and radiation fields at $t=10^3$ years after the impulsive injection by an accelerator, when $D_0=$ 0.1 (blue lines), 0.01 (orange lines), and 0.001 (green lines) $\mathrm{pc^\alpha/yr}$ with $\alpha=$ 1, 1.2, 1.4, 1.6, 1.8, and 2 for subplots from top to bottom and from left to right, respectively. In each subplot, the dashed, dotdashed, and dotted lines show the fluxes of $\pi^0$-decay, IC, and bremsstrahlung $\gamma$ rays, respectively, while the solid lines show the total fluxes. We can see that the fluxes of $\pi^0$-decay $\gamma$ rays dominate over those of IC and bremsstrahlung $\gamma$ rays, and  the characteristic $\pi^0$-decay bump is prominent \citep{Stecker1971}. When $E_\gamma \gtrsim$ 10 TeV, however, the fluxes of IC $\gamma$ rays (resulting mainly from IC scattering off CMB photons) dominate over that of $\pi^0$-decay $\gamma$ rays, while the bremsstrahlung $\gamma$-ray fluxes are generally subdominant to IC and $\pi^0$-decay $\gamma$-ray ones as $E_\gamma >$ 100 MeV. Furthermore, for smaller $D_0$, the $\gamma$-ray spectrum becomes flatter, regardless of the value of $\alpha$. On the other hand, for the same $D_0$, the $\gamma$-ray spectrum becomes flatter, when $\alpha$ is larger.

\begin{figure*}
    \centering
    \includegraphics[width=0.99\textwidth]{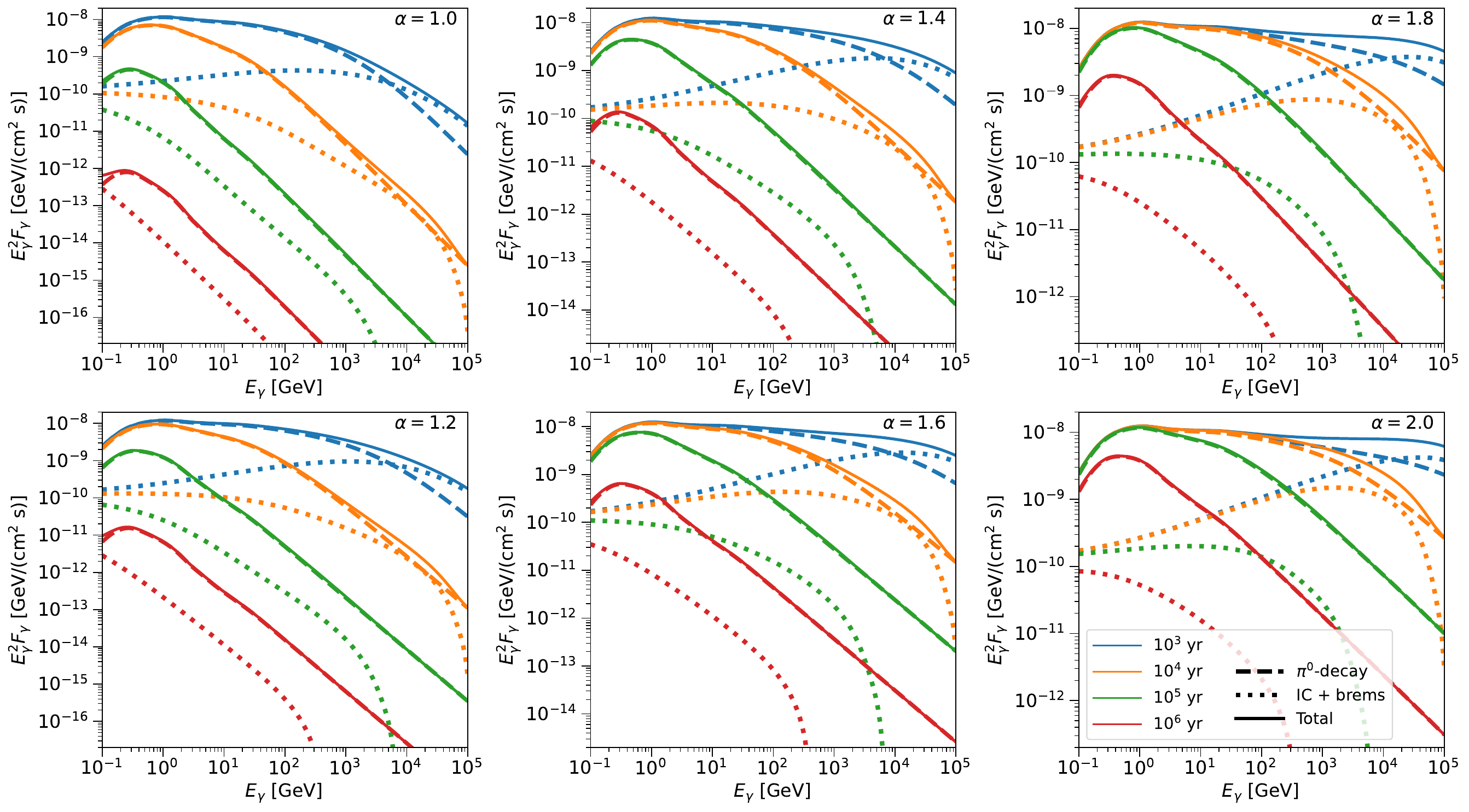}
    \caption{The $\gamma$-ray fluxes at $t=10^3$ (blue lines), $10^4$ (orange lines), $10^5$ (green lines), and $10^6$ (red lines) years after the impulsive injection by an accelerator, when $D_0=0.001~\mathrm{pc^\alpha/pc}$ with $\alpha=$ 1, 1.2, 1.4, 1.6, 1.8, and 2 in each subplot from top to bottom and from left to right. The distance to the accelerator is $d=$ 1 kpc, and the dimension of $\gamma$-ray emission volume is $R=$ 100 pc. The total kinetic energy injected into CR protons is $W_\mathrm{p}=10^{50}~\mathrm{erg}$, the electron-to-proton ratio $K_\mathrm{ep} = 0.001$, the injection spectral index $s=2.2$ for both CR protons and electrons, and the diffusion exponent $\delta=0.5$. The ambient gas density, magnetic field strength, and ISRFs surrounding the accelerator are the same as those specified in discussing Fig. \ref{fig:cooltime}. In each subplot, the dashed and dotted lines show the fluxes of $\pi^0$-decay and IC plus bremsstrahlung $\gamma$ rays, respectively, while the solid lines show the total fluxes.}
    \label{fig:gspec2-impulsive}
\end{figure*}

Fig. \ref{fig:gspec2-impulsive} shows the $\gamma$-ray fluxes at $t=10^3$ (blue lines), $10^4$ (orange lines), $10^5$ (green lines), and $10^6$ (red lines) years after the impulsive injection by an accelerator, when $D_0=0.001~\mathrm{pc^\alpha/yr}$ with $\alpha=$ 1, 1.2, 1.4, 1.6, 1.8, and 2 for subplots from top to bottom and from left to right, respectively. In each subplot, the dashed and dotted lines show the fluxes of $\pi^0$-decay and IC plus bremsstrahlung $\gamma$ rays, respectively, while the solid lines show the total fluxes. Since CR electrons suffer severe energy losses, we can see that the contribution of leptonic processes (namely IC and bremsstrahlung) to the $\gamma$-ray emission becomes increasingly subdominant when $t$ is larger, regardless of the value of $\alpha$. Besides, at high energies, the leptonic $\gamma$-ray spectrum cuts off due to the sharp break existing in the electron spectrum. Thus, as the accelerator becomes more evolved, the surrounding $\gamma$-ray emission becomes increasingly dominated by hadronic processes, however, the $\gamma$-ray emission becomes fainter and its spectrum becomes steeper since the higher energy CRs are escaping more quickly from the emission volume.

\begin{figure*}
    \centering
    \includegraphics[width=0.8\textwidth]{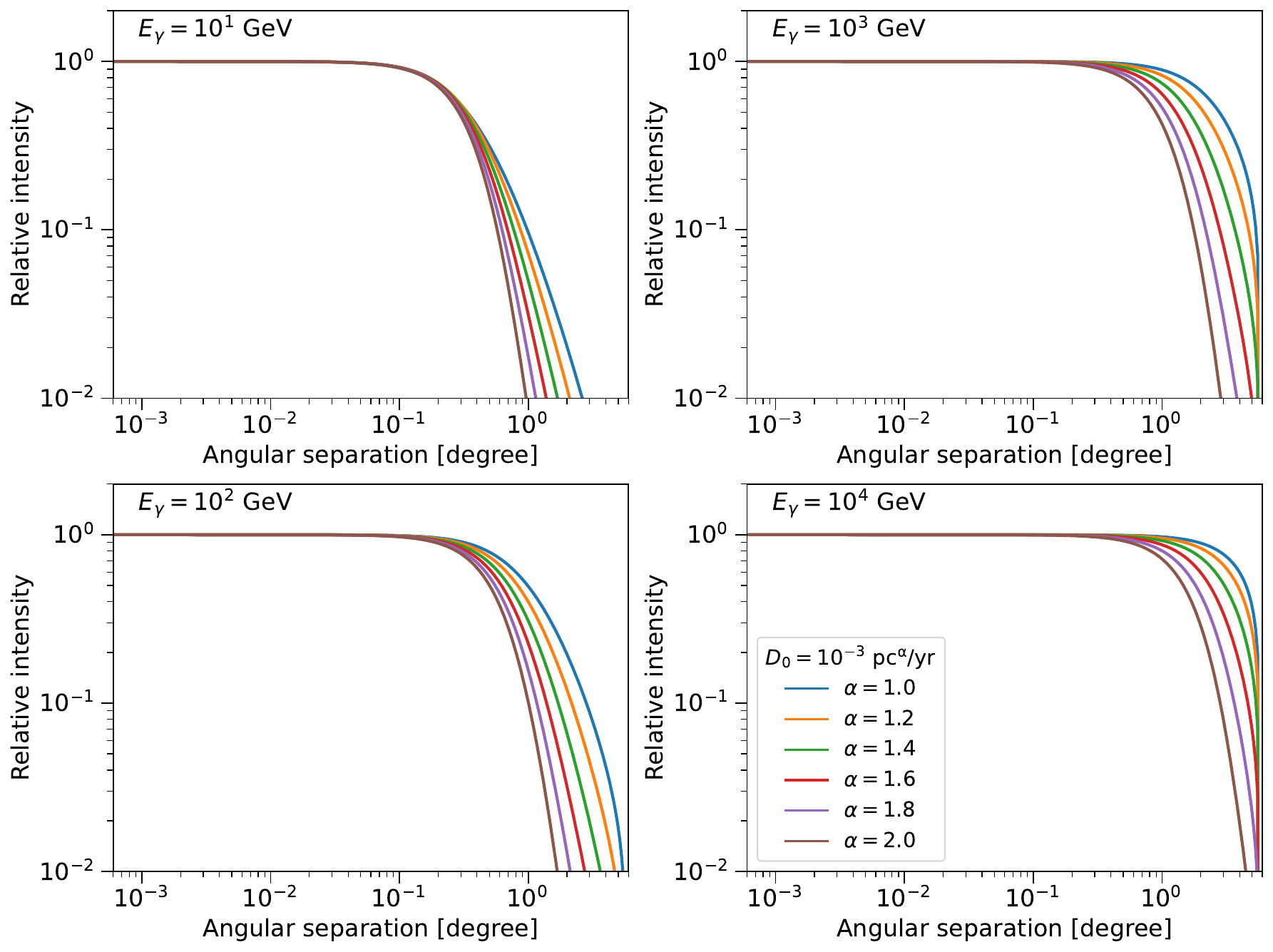}
    \caption{The relative intensity radial profiles of $\gamma$ rays with $E_\gamma=$ 10 (top left panel), 100 (bottom left panel), 1000 (top right panel), and 10000 (bottom right panel) GeV at $t=10^3$ years after the impulsive injection by an accelerator, when $D_0=0.001~\mathrm{pc^\alpha/yr}$ with $\alpha=$ 1 (blue lines), 1.2 (orange lines), 1.4 (green lines), 1.6 (red lines), 1.8 (purple lines), and 2 (brown lines). The distance to the accelerator is $d=$ 1 kpc, and the dimension of $\gamma$-ray emission volume is $R=$ 100 pc. The total kinetic energy injected into CR protons is $W_\mathrm{p}=10^{50}~\mathrm{erg}$, the electron-to-proton ratio $K_\mathrm{ep} = 0.001$, the injection spectral index $s=2.2$ for both CR protons and electrons, and the diffusion exponent $\delta=0.5$. The ambient gas density, magnetic field strength, and ISRFs surrounding the accelerator are the same as those specified in discussing Fig. \ref{fig:cooltime}.}
    \label{fig:gprof-impulsive}
\end{figure*}

Fig. \ref{fig:gprof-impulsive} shows the relative intensity radial profiles of $\gamma$ rays with $E_\gamma=$ 10 (top left panel), 100 (bottom left panel), 1000 (top right panel), and 10000 (bottom right panel) GeV at $t=10^3$ years after the impulsive injection by an accelerator, when $D_0=0.001~\mathrm{pc^\alpha/yr}$ with $\alpha=$ 1 (blue lines), 1.2 (orange lines), 1.4 (green lines), 1.6 (red lines), 1.8 (purple lines), and 2 (brown lines). The relative intensity is defined by $S_\gamma(E_\gamma, \vartheta) = I_\gamma(E_\gamma, \vartheta)/I_\gamma(E_\gamma, \vartheta_0)$,  where $\vartheta$ is the angular separation from the accelerator, and $\vartheta_0 = \arcsin(0.01~\mathrm{pc}/d)$. For each value of $\alpha$, we can see that the radial profile is flat for a given $E_\gamma$, except at the outskirts of the emission volume.  Furthermore, because of the energy-dependent diffusion, the $\gamma$-ray emission is more spatially extended for higher $E_\gamma$. Finally, it is worth noting that the $\gamma$-ray emission is more spatially extended for smaller $\alpha$ at a fixed $E_\gamma$.

\begin{figure*}
    \centering
    \includegraphics[width=0.99\textwidth]{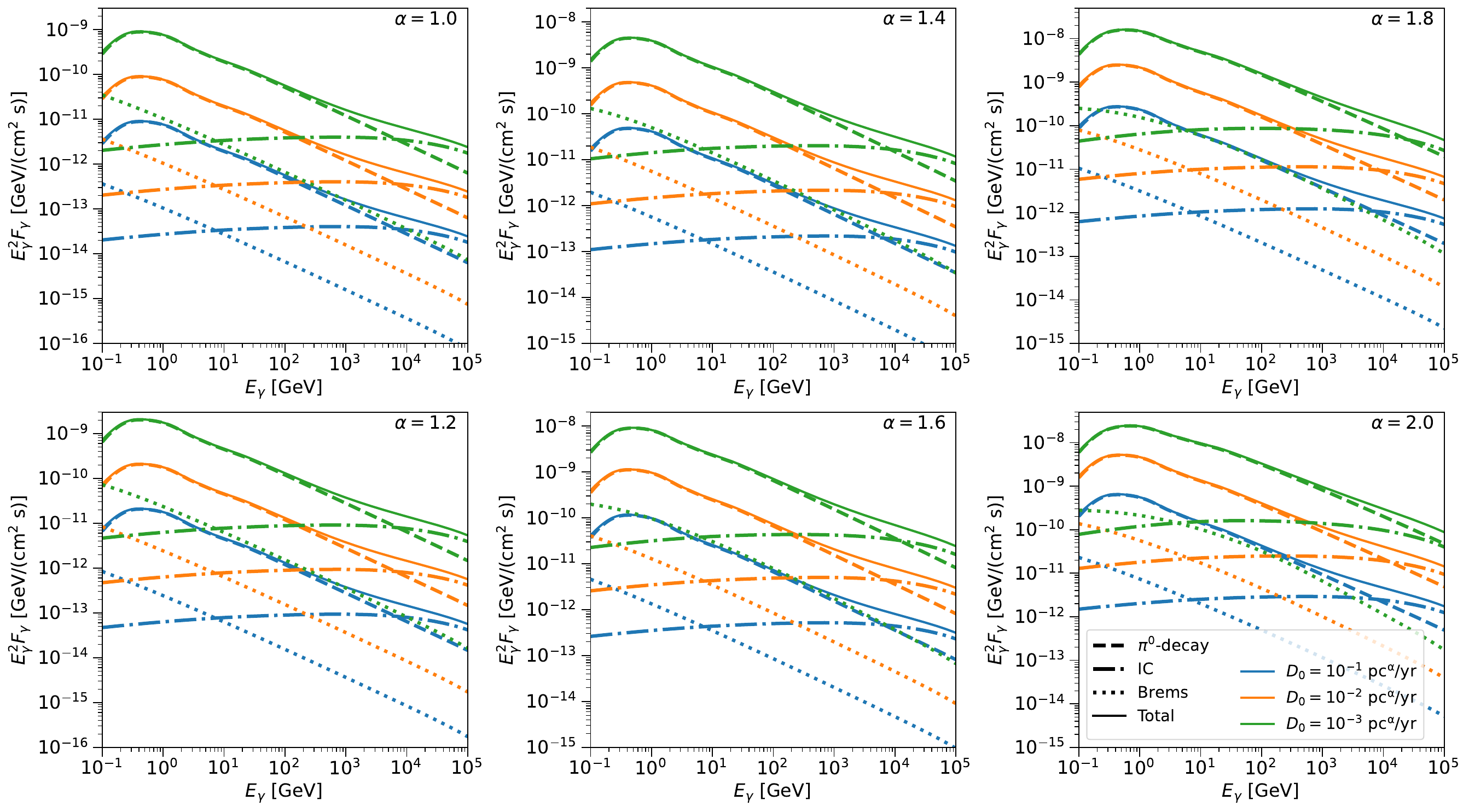}
    \caption{The $\gamma$-ray fluxes at $t=10^6$ years after the stationary injection starting from $t=0$ by an accelerator, when  $D_0=$ 0.1 (blue lines), 0.01 (orange lines), and 0.001 (green lines) $\mathrm{pc^\alpha/yr}$ with $\alpha=$ 1, 1.2, 1.4, 1.6, 1.8, and 2 in each subplot from top to bottom and from left to right. The distance to the accelerator is $d=$ 1 kpc, and the dimension of $\gamma$-ray emission volume is $R=$ 100 pc. The total kinetic power injected into CR protons is $\dot{W}_\mathrm{p}=10^{37}~\mathrm{erg/s}$, the electron-to-proton ratio $K_\mathrm{ep} = 0.001$, the injection spectral index $s=2.2$ for both CR protons and electrons, and the diffusion exponent $\delta=0.5$. The ambient gas density, magnetic field strength, and ISRFs surrounding the accelerator are the same as those specified in discussing Fig. \ref{fig:cooltime}. In each subplots, the dashed, dotdashed, and dotted lines show the fluxes of $\gamma$ rays generated from the decay of $\pi^0$-mesons, IC, and bremsstrahlung, respectively, while the solid lines show the total fluxes.}
    \label{fig:gspec1-stationary}
\end{figure*}

Fig. \ref{fig:gspec1-stationary} shows the $\gamma$-ray fluxes at $t=10^6$ years after the stationary injection starting from $t=0$ by an accelerator, when $D_0=$ 0.1 (blue lines), 0.01 (orange lines), and 0.001 (green lines) $\mathrm{pc^\alpha/yr}$ with $\alpha=$ 1, 1.2, 1.4, 1.6, 1.8, and 2 for subplots from top to bottom and from left to right, respectively. In each subplot, the dashed, dotdashed, and dotted lines show the fluxes of $\pi^0$-decay, IC, and bremsstrahlung $\gamma$ rays, respectively, while the solid lines show the total fluxes. Similar to the impulsive injection, we can see that the fluxes of $\pi^0$-decay $\gamma$ rays dominate over those of IC and bremsstrahlung $\gamma$ rays, and the characteristic $\pi^0$-decay bump is prominent. When $E_\gamma \gtrsim$ 10 TeV, however, the fluxes of IC $\gamma$ rays (resulting mainly from IC scattering off CMB photons) dominate over that of $\pi^0$-decay $\gamma$ rays. Furthermore, for larger $D_0$, the $\gamma$-ray flux becomes smaller, regardless of the value of $\alpha$. However, the spectral shape depends weakly on both $D_0$ and $\alpha$.

\begin{figure*}
    \centering
    \includegraphics[width=0.99\textwidth]{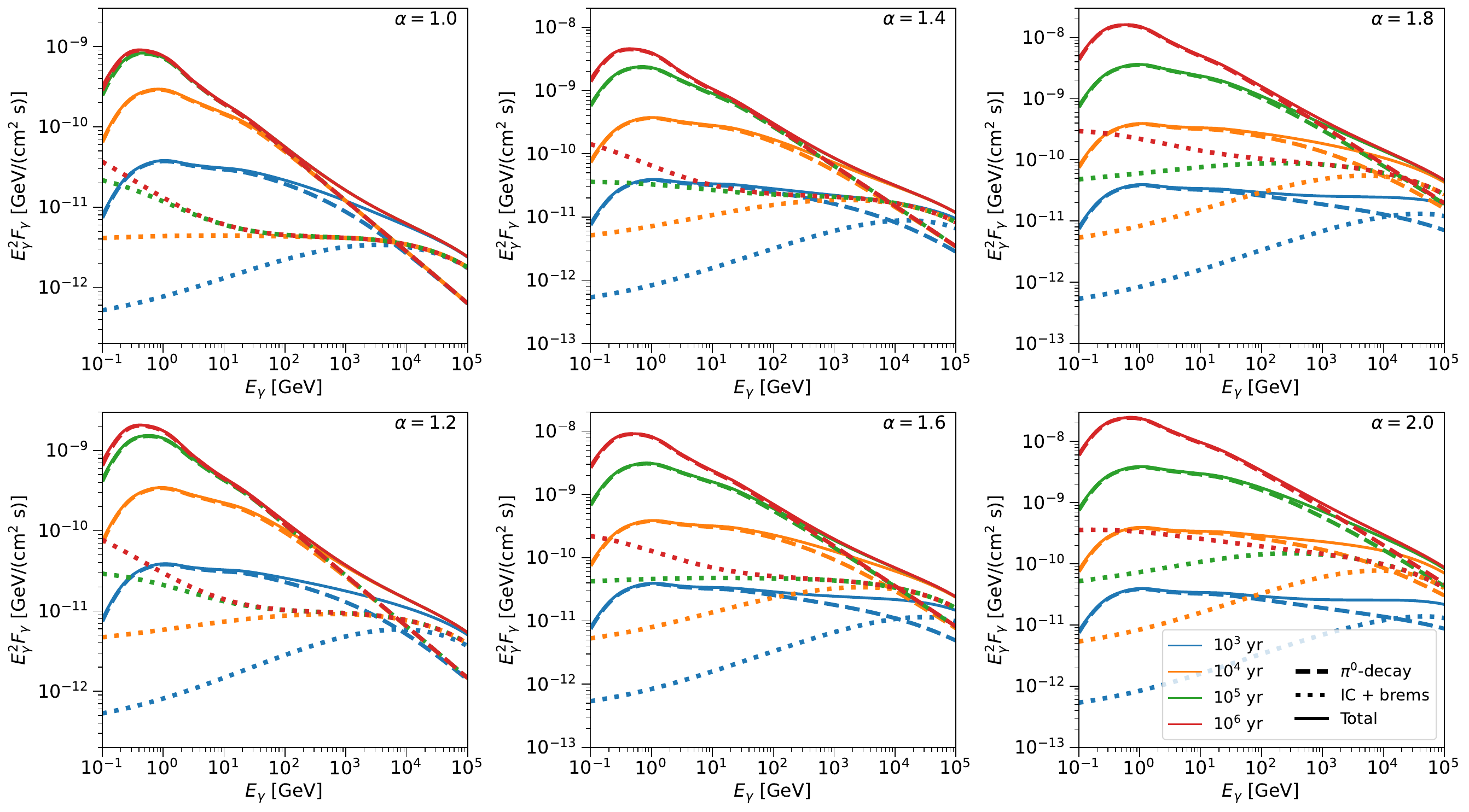}
    \caption{The $\gamma$-ray fluxes at $t=10^3$ (blue lines), $10^4$ (orange lines), $10^5$ (green lines), and $10^6$ (red lines) years after the stationary injection starting from $t=0$ by an accelerator, when $D_0=0.001~\mathrm{pc^\alpha/pc}$ with $\alpha=$ 1, 1.2, 1.4, 1.6, 1.8, and 2 in each subplot from top to bottom and from left to right. The distance to the accelerator is $d=$ 1 kpc, and the dimension of $\gamma$-ray emission volume is $R=$ 100 pc. The total kinetic power injected into CR protons is $\dot{W}_\mathrm{p}=10^{37}~\mathrm{erg/s}$, the electron-to-proton ratio $K_\mathrm{ep} = 0.001$, the injection spectral index $s=2.2$ for both CR protons and electrons, and the diffusion exponent $\delta=0.5$. The ambient gas density, magnetic field strength, and ISRFs surrounding the accelerator are the same as those specified in discussing Fig. \ref{fig:cooltime}. In each subplot, the dashed and dotted lines show the fluxes of $\pi^0$-decay and IC plus bremsstrahlung $\gamma$ rays, respectively, while the solid lines show the total fluxes.}
    \label{fig:gspec2-stationary}
\end{figure*}

Fig. \ref{fig:gspec2-stationary} shows the $\gamma$-ray fluxes at $t=10^3$ (blue lines), $10^4$ (orange lines), $10^5$ (green lines), and $10^6$ (red lines) years after the stationary injection starting from $t=0$ by an accelerator, when $D_0=0.001~\mathrm{pc^\alpha/yr}$ with $\alpha=$ 1, 1.2, 1.4, 1.6, 1.8, and 2 for subplots from top to bottom and from left to right, respectively. In each subplot, the dashed and dotted lines show the fluxes of $\pi^0$-decay and IC plus bremsstrahlung $\gamma$ rays, respectively, while the solid lines show the total fluxes. As the accelerator becomes more evolved, the surrounding $\gamma$-ray emission becomes brighter and its spectra becomes steeper since more CRs with lower energies populate the emission volume.

\begin{figure*}
    \centering
    \includegraphics[width=0.8\textwidth]{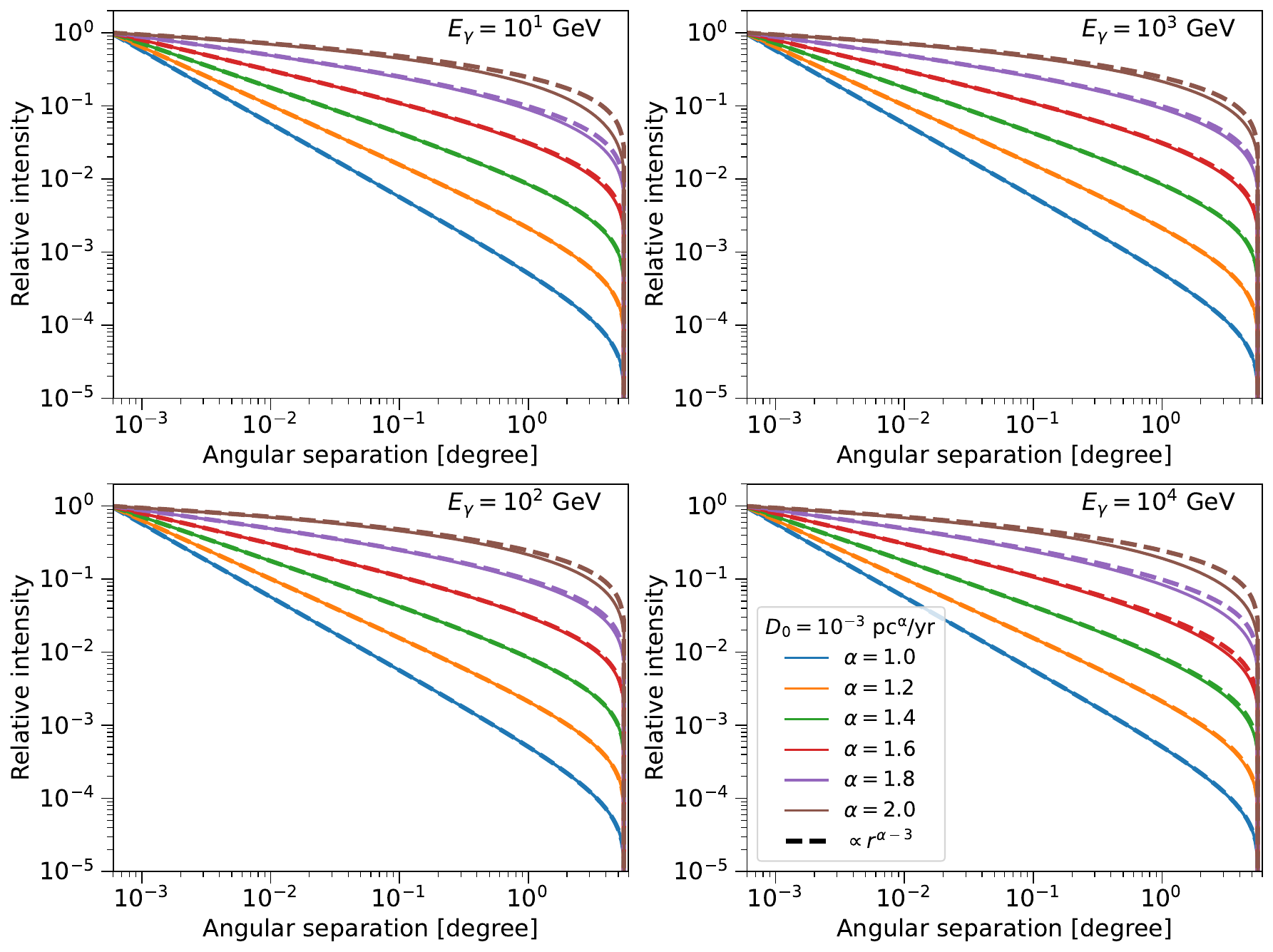}
    \caption{The relative intensity radial profiles of $\gamma$ rays with $E_\gamma=$ 10 (top left panel), 100 (bottom left panel), 1000 (top right panel), and 10000 (bottom right panel) GeV at $t=10^6$ years after the stationary injection starting from $t=0$ by an accelerator, when $D_0=0.001~\mathrm{pc^\alpha/yr}$ with $\alpha=$ 1 (blue lines), 1.2 (orange lines), 1.4 (green lines), 1.6 (red lines), 1.8 (purple lines), and 2 (brown lines). The distance to the accelerator is $d=$ 1 kpc, and the dimension of $\gamma$-ray emission volume is $R=$ 100 pc. The total kinetic power injected into CR protons is $W_\mathrm{p}=10^{37}~\mathrm{erg/s}$, the electron-to-proton ratio $K_\mathrm{ep} = 0.001$, the injection spectral index $s=2.2$ for both CR protons and electrons, and the diffusion exponent $\delta=0.5$. The ambient gas density, magnetic field strength, and ISRFs surrounding the accelerator are the same as those specified in discussing Fig. \ref{fig:cooltime}. The dashed lines show the radial profiles of $\gamma$ rays as defined by Eq. \eqref{eq:plprof}, assuming that the spatial distribution of CRs is proportional to $r^{\alpha - 3}$.}
    \label{fig:gprof-stationary}
\end{figure*}

Fig. \ref{fig:gprof-stationary} shows the relative intensity radial profiles of $\gamma$ rays with $E_\gamma=$ 10 (top left panel), 100 (bottom left panel), 1000 (top right panel), and 10000 (bottom right panel) GeV at $t=10^6$ years after the stationary injection starting from $t=0$ by an accelerator, when $D_0=0.001~\mathrm{pc^\alpha/yr}$ with $\alpha=$ 1 (blue lines), 1.2 (orange lines), 1.4 (green lines), 1.6 (red lines), 1.8 (purple lines), and 2 (brown lines). The relative intensity has the same definition as that for the impulsive injection. As discussed in Sec. \ref{sec:crstationary}, when the cooling time of CRs is larger than the age of the accelerator and the diffusion time scale, their spatial distribution is proportional to $r^{\alpha - 3}$ as $r \ll r_\mathrm{d}$. Assuming that the spatial distribution of CRs is proportional to $r^{\alpha - 3}$, the spatial distribution of produced $\gamma$ rays will be the following

\begin{equation} \label{eq:plprof}
\begin{aligned}
  S_\gamma(\vartheta) &\propto 2\int_0^{\sqrt{R^2 - d^2\sin^2\vartheta}} (l^2 + d^2\sin^2\vartheta)^{\frac{\alpha - 3}{2}} dl \\ &= 2\int_{d\sin\vartheta}^R \frac{r^{\alpha - 2}dr}{\sqrt{r^2 - d^2\sin^2\vartheta}},
\end{aligned}
\end{equation}

\noindent where $d$ is the distance to the accelerator, $R$ is the dimension of $\gamma$-ray emission volume, and $\vartheta$ is the angular separation from the accelerator. In particular, when $\alpha = 1$,

\begin{equation}
  S_\gamma(\vartheta) \propto \frac{4}{d\sin\vartheta}\left\{\arctan\left[\frac{R}{d\sin\vartheta} + \sqrt{\left(\frac{R}{d\sin\vartheta}\right)^2 - 1}\right] - \frac{\pi}{4}\right\},
\end{equation}

\noindent and when $\alpha=2$, 

\begin{equation}
  S_\gamma(\vartheta) \propto 2\ln\left[\frac{R}{d\sin\vartheta} + \sqrt{\left(\frac{R}{d\sin\vartheta}\right)^2-1}\right].
\end{equation}

\begin{figure*}
    \centering
    \includegraphics[width=0.8\textwidth]{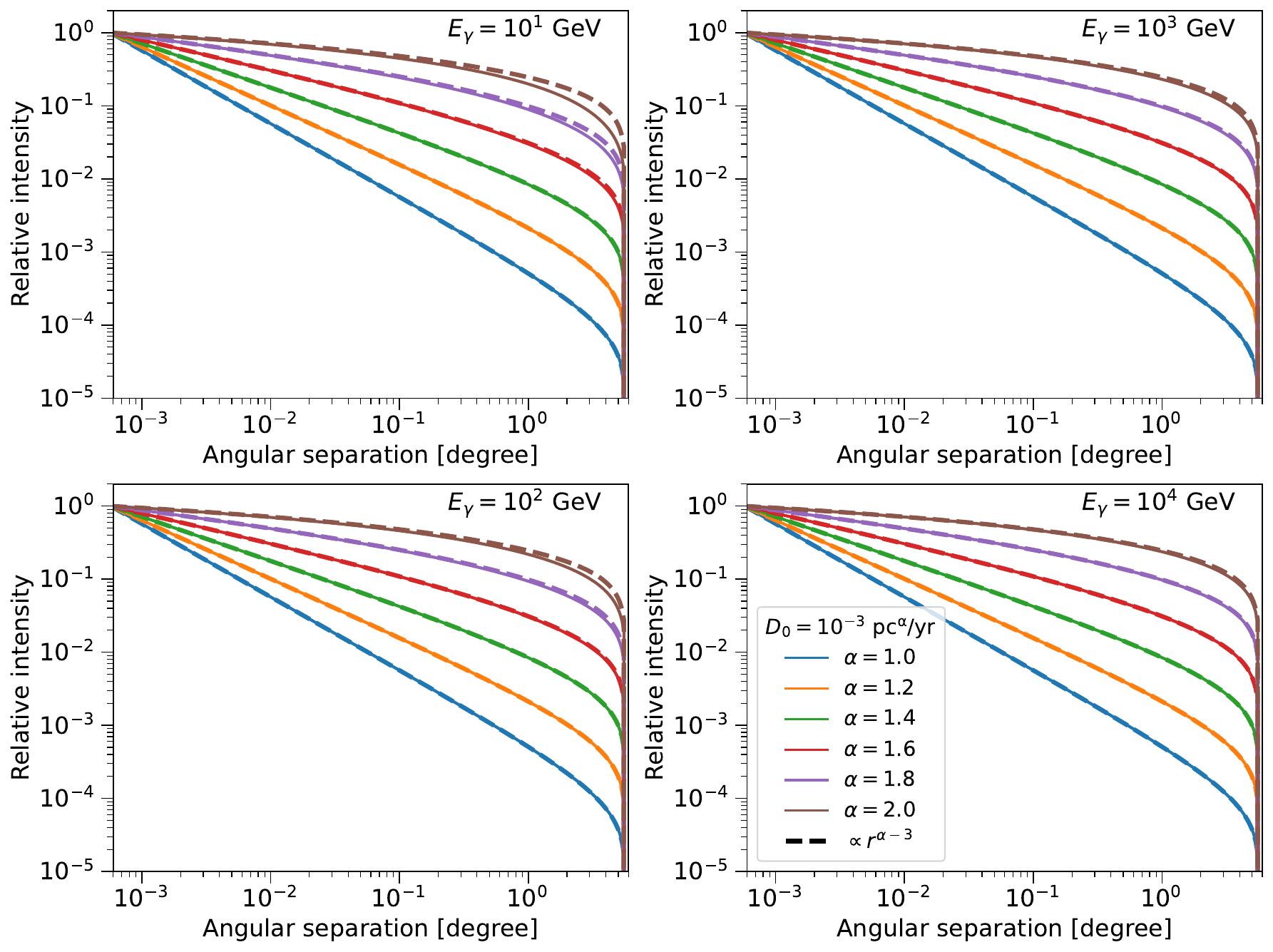}
    \caption{Same as Fig. \ref{fig:gprof-stationary} but only for hadronic ($\pi^0$-decay) $\gamma$ rays.}
    \label{fig:gprof-stationary-hadronic}
\end{figure*}

\begin{figure*}
    \centering
    \includegraphics[width=0.8\textwidth]{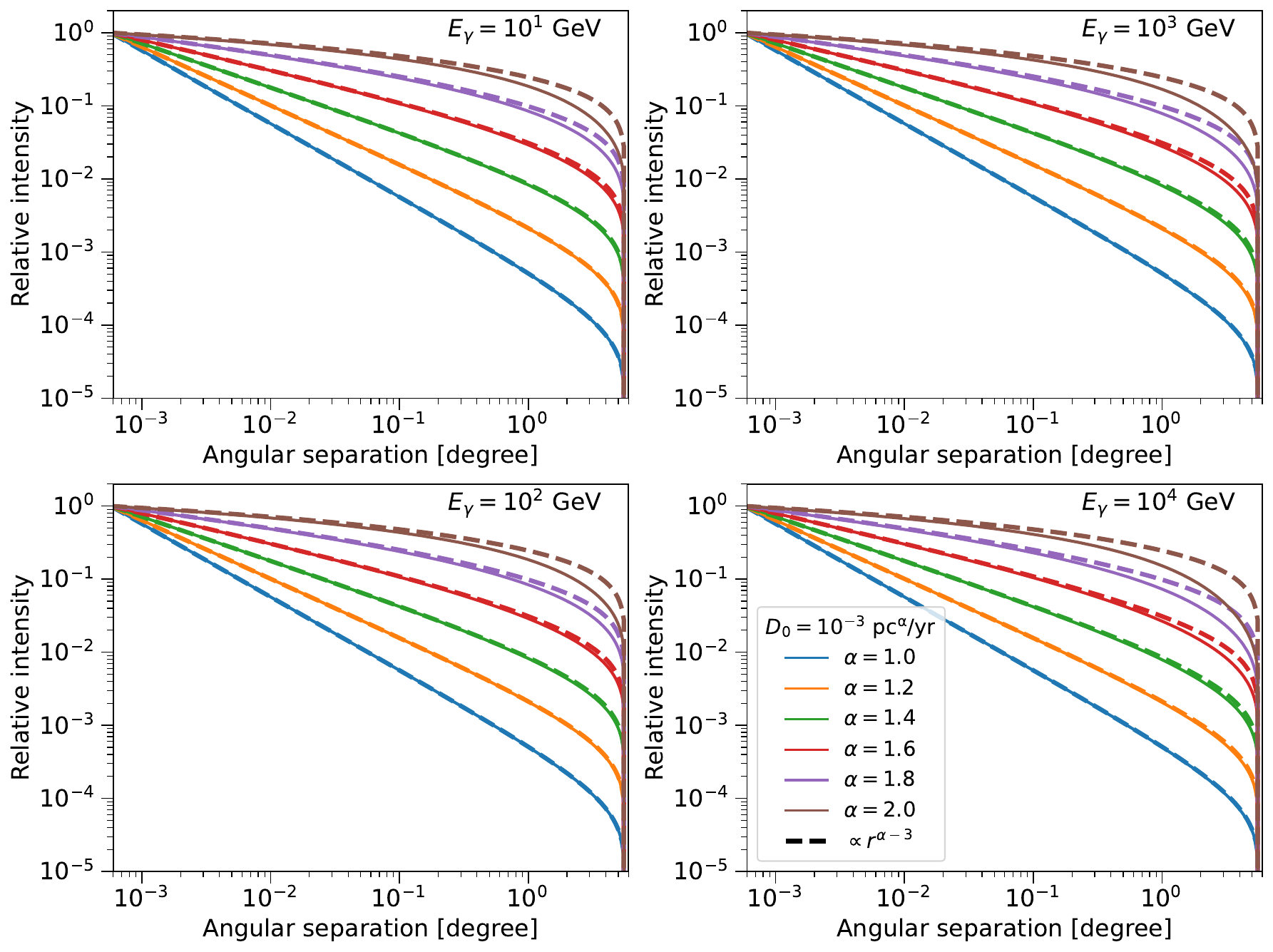}
    \caption{Same as Fig. \ref{fig:gprof-stationary} but only for leptonic (IC plus bremsstrahlung) $\gamma$ rays.}
    \label{fig:gprof-stationary-leptonic}
\end{figure*}

\noindent In Fig. \ref{fig:gprof-stationary}, the dashed lines show the radial profiles as defined by Eq. \eqref{eq:plprof}. For each energy $E_\gamma$, we can see that the radial profile of $\gamma$ rays is in good agreement with that given by Eq. \eqref{eq:plprof}, particularly for smaller $\alpha$. Due to the energy-dependent diffusion, however, the spatial distribution of CRs with lower energies does not follow the $r^{\alpha-3}$-distribution. Moreover, due to the severe energy losses, the spatial distribution of CR electrons with higher energies does not follow the $r^{\alpha-3}$-distribution, too. Therefore, as $E_\gamma$ increases, the difference between the radial profile of $\gamma$ rays and that given by Eq. \eqref{eq:plprof} initially decreases and then increases, for larger $\alpha$ such as $\alpha=$ 1.8 and 2. However, in any case, the difference is not significant. To make it clearer, we also show the radial profiles of hadronic ($\pi^0$-decay) and leptonic (IC plus bremsstrahlung) $\gamma$ rays in Figs. \ref{fig:gprof-stationary-hadronic} and \ref{fig:gprof-stationary-leptonic}, respectively.

%%%%%%%%%%%%%%%%%%%%%%%%%%%%%%%%%%%%%%%
%%%%%%%%%%%%%%%%%%%%%%%%%%%%%%%%%%%%%%%

%%%%%%%%%%%%%%%%%%%%%%%%%%%%%%%%%%%%%%%%%
%%%%% summary %%%%%%%%%%%%%%%%%%%%%%%%%%%%%%%
%%%%%%%%%%%%%%%%%%%%%%%%%%%%%%%%%%%%%%%%%
\section{Summary} \label{sec:summary}

In this article, we studied the propagation of CR protons and electrons in the vicinity of their accelerators. Beyond the normal diffusion, we have assumed that the transport of CRs is described more generally by the superdiffusion, besides taking the energy losses into account. Assuming that the diffusion is momentum-dependent, we investigated the spectral and spatial distributions of CRs at different ages via taking different $D_0$ for both impulsive and stationary injections. In particular, we find that the superdiffusivity impacts significantly on the distributions of CRs, including spectral and spatial ones. When the cooling time scale is larger than the age, $t$, of the accelerator and diffusion time scale, the distribution of CRs is determined by the diffusion length $r_\mathrm{d} = [D(p)t]^{1/\alpha}$. We find that the radial profile tends to being constant for impulsive injection, and it tends to being proportional to $r^{\alpha - 3}$ for stationary injection. In Table \ref{tab:sum}, we summarize the asymptotic spectral index and dependence of radial profile on $r$ after propagation as $r\ll r_\mathrm{d}$ and $r\gg r_\mathrm{d}$, for impulsive and stationary injections.

We also studied the distribution of $\gamma$ rays, produced by the interactions of CRs with ambient gas and radiation fields, within 100 pc around the accelerators, assuming that both gas and radiation fields are homogeneous. We find that superdiffusion impacts significantly the morphology of $\gamma$-ray emission. For impulsive injection, the $\gamma$-ray emission behaves a constant radial profile at the central part, and is more extended for smaller $\alpha$. On the other hand, for stationary injection, the hadronic $\gamma$-ray emission follows $r^{\alpha - 3}$-profile (without integrating along the line of sight) when $E_\gamma$ is high enough, while the leptonic $\gamma$-ray emission can deviate from $r^{\alpha - 3}$-profile at high energies, due to the severe energy losses suffered by electrons. Therefore, by investigating the morphology of $\gamma$-ray emission surrounding the accelerators, we can extract the information on the distribution of CRs in their vicinity, and thus distinguish the superdiffusion from the normal diffusion. Furthermore, if dense gases, such as molecular clouds, exist in the proximity of accelerators, the $\gamma$-ray emission will be enhanced when CRs reach and penetrate them, and consequently the distribution of CRs around the accelerators can be obtained more easily \citep{Aharonian1996}.

%%%%%%%%%%%%%%%%%%%%%%%%%%%%%%%%%%%%%%%%%%%%%%%%%%%%%
\begin{table*}
    \centering
    \begin{threeparttable}
    \begin{tabular}{c|c|c|c|c}
        \hline
        \hline
        injection & $s_\mathrm{prop}$\tnote{\dag} ($r \gg r_\mathrm{d}$, $\alpha<2$) & $s_\mathrm{prop}$\tnote{\dag} ($r\ll r_\mathrm{d}$) & radial profile ($r\gg r_\mathrm{d}$) & radial profile ($r\ll r_\mathrm{d}$)\\
        \hline
        impulsive & $s - \delta$\tnote{\ddag} & $s + 3\delta/\alpha$ & $r^{-\alpha-3}$ \quad ($\alpha < 2$) & constant\\
        & & & $\mathrm{e}^{-r^2/(2r_\mathrm{d})^2}$ \quad ($\alpha = 2$) & \\
        \hline
        stationary & $s - \delta$\tnote{\ddag} & $s + \delta$ & $r^{-\alpha - 3}$ \quad ($\alpha < 2$) & $r^{\alpha - 3}$\\
        & & & $r^{-2}\mathrm{e}^{-r^2/(2r_\mathrm{d})^2}$ \quad ($\alpha = 2$) & \\
        \hline
    \end{tabular}
    \begin{tablenotes}
    \footnotesize
    \item[\dag] $s_\mathrm{prop}$ is the momentum spectral index after propagation.
    \item[\ddag] Only for ultrarelativistic CRs and $\alpha < 2$. However, for transrelativistic CRs, the momentum spectra $N(r,p) \propto v(p) p^{-(s-\delta)}$, where $v(p)$ is the CR velocity, for $\alpha < 2$.
    \end{tablenotes}
\end{threeparttable}
    \caption{Spectral index and radial profile of CRs after propagation, assuming that the injection spectrum is power-law with index $s$, and the diffusion coefficient is given by Eq. \eqref{eq:diffcoe}.}
    \label{tab:sum}
\end{table*}

\section*{Acknowledgements}

Rui-zhi Yang is supported by the National Natural Science Foundation
of China under grants 12588101, 12393854, and by the natural science funding of Sichuan Province under grant 2025ZNSFSC0065. Rui-zhi Yang gratefully acknowledge the support of Cyrus Chun Ying Tang Foundations and of the studio of Academician Zhao Zhengguo, Deep Space Exploration Laboratory.

%%%%%%%%%%%%%%%%%%%%%%%%%%%%%%%%%%%%%%%%%%%%%%%%%%
\section*{Data Availability}

The data underlying this article will be shared on reasonable request to the corresponding author.



\bibliographystyle{mnras}
\bibliography{main} % if your bibtex file is called example.bib

%\begin{thebibliography}{99}
%\end{thebibliography}

%%%%%%%%%%%%%%%%%%%%%%%%%%%%%%%%%%%%%%%%%%%%%%%%%%

%%%%%%%%%%%%%%%%% APPENDICES %%%%%%%%%%%%%%%%%%%%%

%\appendix



%%%%%%%%%%%%%%%%%%%%%%%%%%%%%%%%%%%%%%%%%%%%%%%%%%

% Don't change these lines
\bsp	% typesetting comment
\label{lastpage}
\end{document}